\documentclass[aps,twocolumn,prx,superscriptaddress,longbibliography]{revtex4-2}
\usepackage{graphicx}
\usepackage{dcolumn}
\usepackage{bm}
\usepackage[T1]{fontenc}
\usepackage[latin9]{inputenc}
\usepackage{textcomp}
\usepackage{amsmath}
\usepackage{mathrsfs}

\usepackage{amssymb}
\usepackage{units}
\usepackage{ulem}
\usepackage{xcolor}
\usepackage{comment}
\usepackage[colorlinks,bookmarks=false,citecolor=blue,linkcolor=blue,urlcolor=blue]{hyperref}
\usepackage{soul}
\newcommand{\mr}[1]{{\color{black} #1 }}

\begin{document}

\title{Bogoliubov excitations driven by thermal lattice phonons in a quantum fluid of light}

\author{Ir\'en\'ee Fr\'erot}
\affiliation{Univ. Grenoble Alpes, CNRS, Grenoble INP, Institut N\'{e}el, 38000 Grenoble, France}
\author{Amit Vashisht}
\affiliation{Univ. Grenoble Alpes, CNRS, LPMMC, 38000 Grenoble, France}
\affiliation{Center for Nonlinear Phenomena and Complex Systems, Universit\'e Libre de Bruxelles, CP 231, Campus Plaine, B-1050 Brussels, Belgium}
\author{Martina Morassi}
\affiliation{Centre de Nanosciences et de Nanotechnologies, CNRS, Universit\'{e} Paris-Sud, Universit\'{e} Paris-Saclay, 91120 Palaiseau, France}
\author{Aristide Lema\^itre}
\affiliation{Centre de Nanosciences et de Nanotechnologies, CNRS, Universit\'{e} Paris-Sud, Universit\'{e} Paris-Saclay, 91120 Palaiseau, France}
\author{Sylvain Ravets}
\affiliation{Centre de Nanosciences et de Nanotechnologies, CNRS, Universit\'{e} Paris-Sud, Universit\'{e} Paris-Saclay, 91120 Palaiseau, France}
\author{Jacqueline Bloch}
\affiliation{Centre de Nanosciences et de Nanotechnologies, CNRS, Universit\'{e} Paris-Sud, Universit\'{e} Paris-Saclay, 91120 Palaiseau, France}
\author{Anna Minguzzi}
\affiliation{Univ. Grenoble Alpes, CNRS, LPMMC, 38000 Grenoble, France}
\author{Maxime Richard}
\affiliation{Univ. Grenoble Alpes, CNRS, Grenoble INP, Institut N\'{e}el, 38000 Grenoble, France}
\affiliation{Majulab International Research Laboratory, French National Centre for Scientifique Research, National University of Singapore, Nanyang Technological University, Sorbonne Universit\'e, Universit\'e C\^ote d'Azur, 117543 Singapore}
\affiliation{Centre for Quantum technologies, National University of Singapore, 117543 Singapore}

\begin{abstract}
\mr{The elementary excitations in weakly interacting quantum fluids have a non-trivial nature which is at the basis of defining quantum phenomena such as superfluidity. These excitations and the physics they lead to have been explored in closed quantum systems at thermal equilibrium both theoretically within the celebrated Bogoliubov framework, and experimentally in quantum fluids of ultracold atoms. Over the past decade, the relevance of Bogoliubov excitations has become essential to understand quantum fluids of interacting photons. Their driven-dissipative character leads to distinct properties with respect to their equilibrium counterparts. For instance, the condensate coupling to the photonic vacuum environment leads to a non-zero generation rate of elementary excitations with many striking implications. In this work, considering that quantum fluids of light are often hosted in solid-state systems, we show within a joint theory-experiment analysis that the vibrations of the crystal constitute another environment that the condensate is fundamentally coupled to. This coupling leads to a unique heat transfer mechanism, resulting in a large generation rate of elementary excitations in typical experimental conditions, and to a fundamental non-zero contribution at vanishing temperatures. Our work provides a complete framework for solid-embedded quantum fluids of light, which is invaluable in view of achieving a regime dominated by photon vacuum fluctuations.}
\end{abstract}

\maketitle
\section{INTRODUCTION}

\begin{figure}[tb]
\includegraphics[width=\columnwidth]{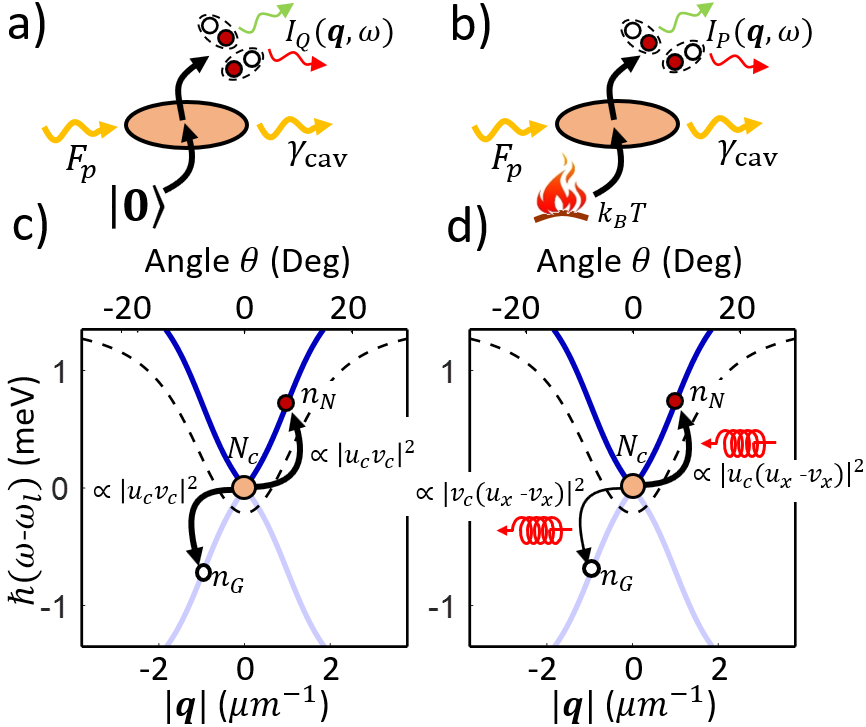}
\caption{Sketch of the two intrinsic mechanisms creating Bogoliubov excitations in a quantum fluid of light embedded in a finite-temperature solid-state micro-resonator. \mr{(a) describes the contribution resulting from the coupling to the extra cavity photon vacuum (noted $|0\rangle$ in the sketch). (b) describes the contribution resulting from the coupling to the bath of lattice phonons at temperature $T$ (symbolized as a fire)}. The experimental configuration considered in this work consists in a steady-state condensate (orange ellipse) involving $\langle N_c\rangle$ photon-like particles (exciton-polaritons) driven at resonance by a laser field of amplitude $F$, and subject to a loss rate $\gamma$. The correlated particle-hole nature of Bogoliubov excitations is shown as a bound red and white symbols. The radiative recombination of the latter is shown as escaping photons. \mr{The photons spectral functions $I_{Q,P}(\textbf{q},\omega)$ are derived in section \ref{section_theory} and given in eq.(\ref{eq_I_quantum}) and eq.(\ref{IpL}) respectively}. (c) and (d) describe the same mechanism as (a) and (b), using the dispersion relation in the framework of the Bogoliubov excitation dispersion relation, which consists of a normal (dark blue) and ghost (light blue) mode, both with a shape that differs from the free polaritons dispersion relation (dashed line). As discussed in the main text, the spontaneous emission rate of Bogoliubov excitations is proportional to $|u_cv_c|^2$ when they result from \mr{coupling to the photon vacuum fluctuations} (c), and to $|u_c(u_x-v_x)|^2$ ($|v_c(u_x-v_x)|^2$) for the normal (ghost) mode, when they result from coupling to thermal lattice phonons (d), where $(u_{c,x},v_{c,x})$ are the coefficients of the Bogoliubov transformation. The subscript $c$ ($x$) refers to the cavity photon (exciton) component of Bogoliubov excitations.}
\label{f1}
\end{figure} 

Interactions between the constituent particles of a quantum fluid play a key role in their response to any space-time perturbations, such as thermal fluctuations, or an obstacle disrupting the quantum flow. They provide a many-body nature to the quantum fluid elementary excitations that results in spectacular macroscopic phenomena, such as superconductivity \cite{Tinkham}, and superfluidity \cite{Kapitza_1938,Allen_1938}. While capturing in full generality many-body excitations represents a daunting theoretical challenge, the celebrated Bogoliubov theory provides a microscopic theoretical framework to describe them in bosonic quantum fluids in the weakly interacting regime, namely when the two-body scattering length is much shorter than the average inter-particle distance. In this regime, the elementary excitations are transformed from free particles (operator $\hat a_{\bf q}$) to correlated particle-hole quasi-particles $\hat \beta_{\bf q}=u_{\bf q} \hat a_{\bf q} + v_{-{\bf q}} \hat a_{-{\bf q}}^\dagger$, where $(u_{\bf q},v_{\bf q})$ are the characteristic Bogoliubov amplitudes, and $\hbar {\bf q}$ is the excitation momentum \cite{Bog_1947,Pit_2003}.  

\mr{This non-trivial nature and its consequences on a quantum fluid behaviour is a strikingly rich research area that has been blossoming, in the context of closed quantum system at thermal equilibrium, for decades. In weakly interacting ultracold atoms, superfluidity \cite{Chikkatur_2000}, and the underlying linear dispersion relation of the elementary excitations \cite{Steinhauer_2002}, as well as the Bogoliubov transformation it stems from \cite{Vogels_2002} have thus been found experimentally in the early 2000s. Since then, the role of Bogoliubov excitations has been investigated and clarified in increasingly complex phenomena such as decoherence \cite{Polkovnikov_2006}, thermalization \cite{Gring_2012,Langen_2013} and revival \cite{Rauer_2018}, or in quantum fluids lacking long range order \cite{Christodoulou_2021}.}






\mr{Over the past decade, the relevance of Bogoliubov excitations in the description of quantum fluids of light has become increasingly evident \cite{IC_2013}. Their necessity emerges from the weakly interacting character of the photons within the fluid, from which a Bogoliubov theory can be rigorously derived \cite{IC_2004}. This strong similarity with their closed equilibrium counterparts allows understanding the emergence of superfluidity in quantum fluids of light \cite{Amo_2009,Fontaine_2018,Michel_2018} as a result of the sound-like dispersion relation of the elementary excitation \cite{IC_2004,Chiao_1999,Kohnle_2011,Claude_2022}. Interestingly the driven-dissipative character of these quantum fluids also leads to a wealth of unique properties. For instance, unlike in closed system, the system openness means that it is coupled to the external world, and in particular to the quantum fluctuations of the photonic vacuum. This coupling leads to a fundamental background generation rate of Bogoliubov excitations on top of the condensate, which is characterized by pairwise quantum correlations \cite{Busch_2014}, with a wide range of applications in dynamical Casimir physics \cite{Koghee_2014}, in Hawking radiation simulators \cite{Gerace_2012,Nguyen_2015,Steinhauer_2016,Jacquet_2020,Jacquet_2020b}, or for the production of quantum states of light \cite{Sarchi_2010,Busch_2014,Regemortel_2018}.}






\mr{In this work, we highlight the fact that in many experimental realizations, the photon vacuum is not the only environment a quantum fluid of light is coupled to. Indeed, for practical purpose, quantum fluids of light are often formed within a solid-state environment such as a nonlinear optical resonators \cite{Chiao_1999,Bolda_2001}, or a polaritonic semiconductor microcavity \cite{Weisbuch_1992,IC_2013}, such that the effective two-body interactions is conveniently provided by dressing the photons with a material electronic transition. In the latter case, photons and bound electron-hole pairs (excitons) are fully hybridized into exciton-polaritons (polaritons), as considered in this work. In the present work, we study both theoretically and experimentally the steady-state flux of Bogoliubov excitations generated on top of a polariton condensate as a result of its coupling to photon vacuum, and to thermal lattice phonons. We show that the solid-state environment, which at first sight could seem anecdotic for photon-like particles, has in fact a profound influence on the condensate properties. Indeed, the crystal lattice vibrations constitute a second full-fledged environment that the condensate is coupled to, and that provides its own contribution of fundamental background excitations on top of the condensate. We show that in our experimental conditions, this contribution not only dominates over that of the photon vacuum fluctuations, but that it also offers a unique opportunity to probe the particle-hole nature of the Bogoliubov excitations. In addition, we evidence that the corresponding non-equilibrium thermal energy flowing between the thermal lattice phonons and the condensate is significantly attenuated by the Bogoliubov transformation. By theoretical extension of our model to $T=0$K, we demonstrate that the most fundamental state of the condensate always involves a contribution of the lattice phonon vacuum, alongside that from the photon vacuum. Finally, we determine a crossover temperature below which the generation of elementary excitations is dominated by the photonic vacuum, which is a necessary condition for the generation of quantum correlations within the condensate excitation spectrum.}

We organize this article as follows; in Section \ref{section_theory}, we develop a Bogoliubov theory of a resonantly driven exciton-polariton condensate, coupled both to lattice phonons and to free space photons. The observables relevant to the experiment are calculated, such as the spectral function of the elementary excitation emission $I(\textbf{q},\omega)$. We report in Section \ref{section_experiment} our measurement of $I(\textbf{q},\omega)$ in a microcavity between temperatures of $6.6\,$K and $12\,$K. By quantitative comparison with the theory, we extract the elementary excitation dispersion relation (Section \ref{sec_dispersion_relations_exp}) and the Bogoliubov transformation amplitudes 
$(u_{\bf q},v_{-{\bf q}})$ (Section \ref{section_bogoliubov_amplitudes}). In the discussion Section \ref{section_discussion} we estimate the experimentally achieved Bogoliubov-transformation induced decoupling from the phonon bath {and} derive a crossover temperature below which \mr{the photon vacuum fluctuations} are expected to overcome the lattice phonons fluctuations. We discuss how to tune the microcavity parameters to achieve a refined control over both phenomena. Finally, section \ref{section_conclusion} offers some concluding remarks.

\section{Theory}
\label{section_theory}
\subsection{Microscopic model and observables}

In this work, we investigate a quantum fluid of light consisting of resonantly driven exciton-polaritons \cite{Weisbuch_1992,IC_2013} (henceforth denoted as 'polaritons'), hybrid quasi-particles obtained when photons confined in a cavity are in the strong coupling regime with an excitonic transition (bound electron-hole pairs) provided by a semiconductor planar quantum well. Their excitonic component provides them with two-body interactions, as well as interactions with the bath of acoustic solid-state lattice phonons, while  the photonic fraction mediates the coupling to the resonant laser drive and to the extra-cavity free propagating photons that constitute the measured observable. The Hamiltonian describing all these interactions expressed in the exciton-photon basis thus consists of the following contributions:
\begin{equation}
    \hat{\cal H} = \hat{\cal H}_{\rm pol}^{(0)} 
    + \hat {\cal V}_{xx} + \hat {\cal V}_{\rm sat}
    + \hat{\cal H}^{(0)}_{\rm ph} + \hat{\cal V}_{xp} +\hat{\cal V}_{\rm out},
\label{eq:Ham-sys}
\end{equation}
where 
\begin{eqnarray}
    \hat{\cal H}_{\rm pol}^{(0)} &=& \hbar \sum_{\bf q} \left[ \omega_{x,{\bf q}} \hat b_{\bf q}^\dagger \hat b_{\bf q}
    + \omega_{c,{\bf q}} \hat a_{{\bf q}}^\dagger \hat a_{{\bf q}}
    + \frac{\Omega}{2} (\hat a_{\bf q}^\dagger \hat b_{\bf q} + {\rm h.c.})
    \right] \nonumber\\
    &&+ \left(f_p(t)^* \hat a_{{\bf q}_p} + {\rm h.c.}\right)
    \label{eq_H_pol}
\end{eqnarray}
describes the interaction between cavity photons (operator $\hat a_{\bf q}$) and quantum well excitons ($\hat b_{\bf q}$ ) of in-plane momentum ${\bf q}$, in which the strong coupling regime is described by the third term in the sum, with $\hbar\Omega$ being the Rabi splitting \cite{Deveaud_2007,Kavokin_2007} separating the upper and lower polariton modes when $\hat{\cal H}_{\rm pol}^{(0)}$ is in its diagonal form \cite{Hopfield_1958}. The last term in $\hat{\cal H}_{\rm pol}^{(0)}$ describes the coherent laser drive with $f_p(t)=F_p e^{i \omega_{\rm las} t} $ of amplitude $F_p$ with corresponding  in-plane momentum $\hbar {\bf q}_p$. 
Concerning the interaction terms, we take
 \begin{equation}
   \hat {\cal V}_{xx} = (\hbar g_x /2) \sum_{{\bf k},{\bf k}',{\bf q}}
   \hat b_{{\bf k}+{\bf q}}^\dagger \hat b_{{\bf k}'-{\bf q}}^\dagger \hat b_{{\bf k}'} \hat b_{\bf k}  .
   \label{H_coulomb}
\end{equation}
This describes the Coulomb-mediated interactions between excitons, of strength $g_x$, that contributes to two-body interactions between polaritons \cite{Ciuti_1998}. 
Furthermore,
\begin{equation}
   \hat {\cal V}_{\rm sat} = (-\hbar g_s/2)\sum_{{\bf k},{\bf k}',{\bf q}}
   (\hat a_{{\bf k}+{\bf q}}^\dagger \hat b_{{\bf k}'-{\bf q}}^\dagger \hat b_{{\bf k}'} \hat b_{\bf k} + {\rm h.c.})
   \label{H_sat}
\end{equation}
describes an additional interaction mechanism between excitons, of strength $g_s$ and often referred to as saturation nonlinearity. It results from the fact that excitons consist of bound fermions, i.e. electrons and holes, so that the creation of an exciton produces a non-zero fermionic phase-space filling, that in turn reduces the photon-creation probability of a second exciton \cite{Rochat_2000}. The term
\begin{equation}
    \hat{\cal H}^{(0)}_{\rm ph} = \hbar\sum_{{\bf q},k_z} \omega_{{\bf q},k_z}^{(\rm ph)} \hat c^\dagger_{{\bf q},k_z} \hat c_{{\bf q},k_z} 
\end{equation}
describes the three dimensional continuum of harmonic \mr{longitudinal lattice vibration modes or longitudinal acoustic (LA) phonons, with bosonic operators $\hat c_{{\bf q},k_z}$. $\omega_{{\bf q},k_z}^{(\rm ph)}=v_s\sqrt{{\bf q}^2 + k_z^2}$ is the phonon dispersion relation, with $v_s$ the LA phonons sound velocity. Transverse acoustic (TA) phonons have a much weaker coupling strength with excitons \cite{Krummheuer_2002} and are thus neglected thereafter.} ${\bf q}=(q_x,q_y)$ is the two-dimensional momentum in the plane of the microcavity spacer and quantum well, and $k_z$ is in the orthogonal direction. \mr{The interaction between excitons and LA phonons is dominated by the elastic deformation potential and reads \cite{Piermarocchi_1996,Savenko_2013}:} 
\begin{equation}
     \hat{\cal V}_{xp} = i\hbar\sum_{{\bf q},k_z} g_{xp}({\bf q},k_z)(\hat c_{{\bf q},k_z} - \hat c_{-{\bf q},k_z}^\dagger) \sum_{{\bf q}'} \hat b^\dagger_{{\bf q}+{\bf q}'} \hat b_{{\bf q}'} ~,
     \label{eq_Vxp}
\end{equation}
where $g_{xp}({\bf q},k_z)$ is the momentum-dependent interaction strength. The detailed expression for $g_{xp}({\bf q},k_z)$ is given is Appendix  \ref{app_gamma_xp}. Finally, the term
\mr{\begin{equation}
   \hat{\cal V}_{\rm out} =  \hbar \sum_{{\bf q},k_z} \left\{ \omega_{{\bf q},k_z}^{(\alpha)} \hat\alpha_{{\bf q},k_z}^\dagger \hat\alpha_{{\bf q},k_z} +  \kappa [\hat\alpha_{{\bf q},k_z}^\dagger \hat a_{\bf q} + {\rm h.c.}]
    \right\} 
    \label{eq_V_photons}
\end{equation}}
describes the conversion of intracavity photons into extracavity free propagating photons, described by the bosonic operator $\hat\alpha_{{\bf q},k_z}$. Cavity photons can tunnel at a rate \mr{$\kappa$  into this continuum across the mirrors, and vice-versa, as a result of the finite reflectivity of the mirrors. $\kappa$ is safely considered constant within the momentum and frequency range of the experiment.} The first term in Eq.~\eqref{eq_V_photons} describes the extra-cavity free propagating photon energy, whose dispersion relation in vacuum is $\omega_{{\bf q},k_z}^{(\alpha)}=c\sqrt{{\bf q}^2 + k_z^2}$ with $c$ the speed of light. The second term describes the tunnel coupling mechanism.

The experimental observable of focus in the current work is the extracavity photon intensity $I({\bf q},\omega)$ resolved both in frequency and momentum. Using an input-output formalism detailed in Appendix \ref{app_input_output} we derive a general relation between the intracavity photon field and the extracavity photon intensity that \mr{simplifies into}
\mr{\begin{equation}
    I({\bf q},\omega) =   \frac{\gamma_{\rm cav}}{\pi} \int_{-\infty}^{+\infty} d\tau  e^{-i\omega \tau} \langle \hat a_{\bf q}^\dagger(\tau) \hat a_{\bf q}(0) \rangle ~,
   \label{eq_output_signal_1}
\end{equation}}
where we have used the fact that the extracavity photons are in a vacuum state (an excellent approximation considering that our photons are in the $\sim 1.5\,$eV energy range), and \mr{considered that the system has reached a steady-state as is the case in the experiment. $\gamma_{\rm cav} \propto \kappa^2$ is the cavity loss rate (see Appendix \ref{app_input_output} for further details).} We thus proceed to derive the two-time correlator $\langle \hat a_{\bf q}^\dagger(\tau) \hat a_{\bf q}(0) \rangle$ in presence of both \mr{photon vacuum fluctuations} and a thermal population of acoustic phonons.

In our experimental configuration, polariton-polariton interactions are relatively weak, i.e. the associated scattering length is much smaller than the interparticle distance, and the driving laser intensity is large enough to induce a macroscopic population of the steady-state excitonic and photonic modes \cite{Pit_2003}. This justifies a mean-field treatment for the condensate and the Bogoliubov approximation for the description of the excitations on top of it.

\subsection{Bogoliubov theory}

\noindent\textit{Mean-field equation for the condensate.--} 
We first need to determine the mean-field steady-state of the system. Writing the Heisenberg equations of motion for the cavity photons and excitons at the laser wavevector ${\bf q}_p$, and setting, as per the mean-field approximation, $\langle \hat b_{{\bf q}_p}\rangle=\psi_x $ and $\langle \hat a_{{\bf q}_p} \rangle = \psi_c$, we find the steady-state equations:
\begin{eqnarray}
    &(&\omega_x-i\gamma_x/2+g_x|\psi_x|^2)\psi_x+\nonumber\\
    &(&\Omega/2-g_s|\psi_x|^2)\psi_c-g_s\psi_x^2\psi_c^*=0
\label{eq_Itp1}
\end{eqnarray}
and:
\begin{eqnarray}
    &(&\omega_c-i\gamma_{\rm cav}/2)\psi_c + \nonumber \\
    &(&\Omega/2-g_s|\psi_x|^2/2)\psi_x+F_p=0~,
\label{eq_Itp2}
\end{eqnarray}
where $n_{x,c}=|\psi_{x,c}|^2$ are the excitonic and photonic densities and from here on the excitonic and photonic frequencies are shifted by the laser frequency $\omega_{\rm las}$. For details on the full solution to Eqs.~(\ref{eq_Itp1}-\ref{eq_Itp2}) see Appendix~\ref{app-meanfield}.

Note that we have introduced a phenomenological decoherence rate for the excitonic transition of the form $\gamma_x({\bf q})=\gamma_{x,0}+\beta q^2$ ($\beta>0$) that describes in an effective way the fact that polaritons of higher energy (and hence of higher $|{\bf q}|$) interact more strongly with the quantum well imperfections (see e.g. \cite{Klembt_2018,Munoz_2019,Scarpelli_2022} for details).
We do not attempt to describe this effect in our model as it would far exceed our scope without clear benefit for the purpose of this work.\\

\noindent\textit{Bogoliubov approximation.--} We then use the Bogoliubov approximation to reduce the interaction terms to a quadratic form. This approximation amounts to: i) assume that both the cavity photon and the exciton modes at the laser momentum ${\bf q}_p$ are macroscopically occupied; and ii) neglect in the Hamiltonian terms involving more than two operators at ${\bf q}\neq {\bf q}_p$. We thus derive the resulting quadratic exciton-exciton interaction terms $\hat{\cal V}_{xx}$ and $\hat{\cal V}_{\rm sat}$ that describe interactions occurring within the condensate, and with final states outside the condensate with momenta ${\bf q}_p+{\bf q}$ and ${\bf q}_p-{\bf q}$ (see Appendix \ref{sec_app_interaction_bogo} for the detailed expression). The phonon-exciton interaction becomes
\begin{eqnarray}
    \hat{\cal V}_{xp} &=& i\hbar\sqrt{n_x} \sum_{{\bf q},k_z} g_{xp}({\bf q}, k_z) \times \nonumber\\ 
    &&(\hat c_{{\bf q},k_z} - \hat c^\dagger_{-{\bf q},k_z})(\hat b_{{\bf q}_p+{\bf q}}^\dagger + \hat b_{{\bf q}_p-{\bf q}}) ~.
    \label{eq_Vxp_bogo}
\end{eqnarray}
and describes the scattering of a condensate exciton via emission or absorption of a phonon into an excited state with momentum ${\bf q}_p + {\bf q}$ or ${\bf q}_p + {\bf q}$. \mr{This term represents a key ingredient in this work as it describes the condensate interaction with the solid-state environment.}

Under the Markov approximation for the damping kernel associated to the bath of extra-cavity photons and the bath of solid-state phonons, the excitations take the following final form (see Appendix \ref{eom-phot}, \ref{eom-exc}, \ref{bath}, \ref{eom-complete} for details):
\begin{equation}
 \hat{\cal{A}}_{{\bf q}_p,{\bf q}}(\omega)=i[\omega \mathbf{1}-M_{\bf q}]^{-1} \hat{\cal{F}}_{{\bf q}_p, {\bf q}} 
    \label{eq_input_output_final}
\end{equation}
where we have set for short-hand notation 
\begin{equation}
\hat{\cal{A}}_{{\bf q}_p,{\bf q}}(\omega)=[\hat a_{{\bf q}_p+{\bf q}}(\omega),\hat b_{{\bf q}_p+{\bf q}}(\omega),\hat a^\dagger_{{\bf q}_p-{\bf q}}(\omega), \hat b^\dagger_{{\bf q}_p-{\bf q}}(\omega)]^T ,
\end{equation}
and introduced the Langevin force vector
\begin{eqnarray}
&\hat{\cal{F}}_{{\bf q}_p,{\bf q}}(\omega)=[&\hat F_{{\bf q}_p+{\bf q}}(\omega),
\hat f_{{\bf q}}(\omega)-\hat f^\dagger_{-{\bf q}}(\omega), \nonumber \\
&&\hat F^\dagger_{{\bf q}_p-{\bf q}}(\omega),f^\dagger_{-{\bf q}}(\omega)- \hat f_{{\bf q}}(\omega)]^T 
\end{eqnarray}
The matrix $M_{\bf q}$ reads:
\begin{widetext}
\begin{equation}
    M_{\bf q} = \begin{pmatrix}
        \omega_{c,{\bf q}_p+{\bf q}} - i \gamma_{\rm cav} &  -2\mu_s +\Omega/2  & 0 & -\mu_s \\ 
        - 2\mu_s+\Omega/2 & \omega_{x,{\bf q}_p+{\bf q}} - i\gamma_{x,{\bf q}_p+{\bf q}} + 2{\rm Re}(\mu_{sx}) & -\mu_s & \mu_{sx} \\
        0 & \mu_s & -\omega_{c,{\bf q}_p-{\bf q}} - i\gamma_{\rm cav}& 2\mu_s  -\Omega/2  \\
        \mu_s & -\mu_{sx}^* &2 \mu_s- \Omega/2 & -\omega_{x,{\bf q}_p-{\bf q}} - i\gamma_{x, {\bf q}_p-{\bf q}} - 2 {\rm Re}(\mu_{sx})
    \end{pmatrix} ~,
    \label{eq_M_mat}
\end{equation}
\end{widetext}
with $ \mu_s =g_s n_x/2$ and $ \mu_{sx} = g_x n_x - g_s \sqrt{n_x n_c} e^{-i\phi}$. 

The Langevin force for cavity photons, that results from their coupling to the extracavity photons reads:
\begin{equation}
\hat F_{\bf q}(t) = -i\sum_{k_z} \kappa_{{\bf q},k_z} \hat{\alpha}_{{\bf q},k_z}^{(in)} e^{-i\omega_{{\bf q},k_z}^{(\alpha)} t} ~, 
\label{eq_langevin_force_photons-m}
\end{equation}
where $\hat\alpha_{{\bf q},k_z}^{(in)} := e^{i\omega_{{\bf q},k_z}^{(\alpha)} t_0}\hat\alpha_{{\bf q},k_z}(t_0)$. As discussed below, this force  \mr{stems from the photon vacuum fluctuations}. Similarly, the Langevin force   acting on excitons, as a result of their coupling to the thermal phonons bath is
\begin{equation}
  \hat f_{\bf q}(t) = \sqrt{n_x} \sum_{k_z} g_{xp}({\bf q},k_z)  \hat{c}_{{\bf q},k_z}^{(in)} e^{-i\omega_{{\bf q},k_z}^{(\rm ph)}t} ~,
  \label{eq_langevin_force_phonons-m}
\end{equation}
where $\hat{c}_{{\bf q},k_z}^{(in)} := \hat{c}_{{\bf q},k_z}(t_0) e^{i\omega_{{\bf q},k_z}^{(\rm ph)}t_0}$.
Similar equations of motion have been derived in the literature \cite{Ciuti_2005,Sarchi_2010}, but the interaction with lattice phonons has not been included so far to the best of our knowledge.\\

\noindent\textit{Bogoliubov eigenmodes.--} Equations \eqref{eq_input_output_final}-\eqref{eq_M_mat} describe how the cavity photons and excitons hybridize due to: (i) the exciton-photon Rabi coupling $\Omega$; and (ii) two-body interactions; as well as their forcing by coupling to lattice phonons, decoherence via the excitonic component, and dissipation into extracavity photons. 
The $M_{\bf q}$ matrix is readily brought onto a diagonal form by $ P_{\bf q} M_{\bf q} P_{\bf q}^{-1}$. \mr{The $P_{\bf q}$ matrix, whose explicit definition is given in Appendix \ref{eom-complete}, contains the $u,v$ coefficients of the Bogoliubov transformation. In particular, the (bosonic) lower-polariton operator reads:
\begin{eqnarray}
    \hat\beta_{lp,c,{\bf q}_p+{\bf q}} &= &u_{lp,c,{\bf q}_p+{\bf q}} \hat a_{{\bf q}_p+{\bf q}} + u_{lp,x,{\bf q}_p+{\bf q}} \hat b_{{\bf q}_p+{\bf q}} + \nonumber\\
    &&v_{lp,c,{\bf q}_p-{\bf q}} \hat a^\dagger_{{\bf q}_p-{\bf q}} + v_{lp,x,{\bf q}_p-{\bf q}} \hat b^\dagger_{{\bf q}_p-{\bf q}} ~.
\end{eqnarray}
}  There are thus four Bogoliubov coefficients per mode in our model instead of the usual two in the Bogoliubov theory. This is because the bosonic quasi-particles that constitute our condensate are exciton-photon hybrids, such that $u_{j,x}$ for instance characterizes both the particle and the excitonic contribution of the Bogoliubov state in mode $j$; $v_{j,c}$ the hole and photonic contributions, and so on. In order to ensure bosonic commutation relations of the new quasiparticle field operators the Bogoliubov coefficients are normalized according to $|u_{c,{\bf q}}|^2 +|u_{x,{\bf q}}|^2 - |v_{c,-{\bf q}}|^2 - |v_{x,-{\bf q}}|^2 = 1$ for all modes. For the sake of compactness, we drop the ${\bf q}_p$ dependence as well as the `lp' subscript for the lower polariton coefficients, that we thus note $(u_{c,{\bf q}},u_{x,{\bf q}},v_{c,-{\bf q}},v_{x,-{\bf q}})$ in the following sections.

\subsection{Dispersion relations}
\label{sec_dispersion_relations_exp}
The eigenvalues of $M_{\bf q}$ provide the four dispersion relations \mr{$\omega_{\{lp,up;N,G\}}({\bf q})$ and damping rates $\gamma_{\{lp,up;N,G\}}({\bf q})$ of the Bogoliubov excitations, where labels 'lp' and 'up' refer to the lower and upper polariton modes, and labels 'N' and 'G' refer to the Normal and Ghost Bogoliubov excitation branches that exists for each polariton modes, and correspond respectively to the positive and negative excitation bands with respect to the condensate energy. Note that} 'lp' and 'up' are still good labels as long as we can neglect the up/lp mixing induced by the Bogoliubov transformation, a condition which is well satisfied as long as $\Omega\gg (\mu_{s},\mu_{xs})$, as is the case in our experiment \cite{Ciuti_2005}. 

Examples of thus obtained lower polariton normal (dark solid line) and ghost (light solid line) mode dispersion relation are shown in Fig.~\ref{f1}(c,d), and compared with the free polariton dispersion relation ($\mu_s=\mu_{sx}=0$, dashed line).

\subsection{Emission intensity $I({\bf q},\omega)$}
In order to determine $I({\bf q},\omega)$ we need to evaluate the intracavity photon correlator in Eq.~\eqref{eq_output_signal_1}. This is achieved by using Eq.~\eqref{eq_input_output_final} and  Fourier transforming back to time domain $\hat a_{\bf q}(t) = \int_{-\infty}^\infty (d\omega/2\pi) e^{-i\omega t} \hat a_{\bf q}(\omega)$. The relevant term in Eq.~\eqref{eq_output_signal_1} is the equation of motion for the intra-cavity photon operator:
\begin{eqnarray}
    \hat a_{{\bf q}_p + {\bf q}}(\omega) &=& 
       G_{11} \hat F_{{\bf q}_p + {\bf q}}(\omega) +
       G_{13} \hat F_{{\bf q}_p - {\bf q}}^\dagger(\omega) \nonumber\\
       &&+(G_{12} - G_{14}) [\hat f_{\bf q}(\omega) - \hat f^\dagger_{-\bf q}(\omega)] ~,
       \label{eq_output_signal_2}
\end{eqnarray}
that involves both the quantum and lattice phonon fluctuations, and where we have set $G({\bf q},\omega) = i[\omega \mathbf{1}-M_{\bf q}]^{-1}$, and  the Langevin forces $\hat F$ and $\hat f$ are given respectively by the Fourier transforms of Eqs.~\eqref{eq_langevin_force_photons-m} and \eqref{eq_langevin_force_phonons-m}. We discuss separately the two different fluctuation contributions.

\subsubsection{\mr{Photon vacuum fluctuations} contribution to $I({\bf q},\omega)$}
We first focus on the contribution from \mr{photon vacuum fluctuations and calculate $\langle \hat a_{\bf q}^\dagger(\tau) \hat a_{\bf q}(0) \rangle$. Note that this contribution has been derived in a similar way in \cite{Busch_2014}}. Given that the initial state for the system is the vacuum for the photon modes, only the $\hat F^\dagger$ term contributes to the output signal \mr{and we find:} 
\mr{\begin{eqnarray}
    \langle \hat a_{{\bf q}_p + {\bf q}}^\dagger(\tau) \hat a_{{\bf q}_p + {\bf q}}(0) \rangle_Q =   \nonumber\\
     \sum_{k_z} \kappa^2 e^{i\tau\omega_{{\bf q}_p - {\bf q},k_z}^{(\alpha)}} |G_{13}({\bf q},\omega_{{\bf q}_p - {\bf q},k_z}^{(\alpha)})|^2 ~.
\end{eqnarray}
From Eq.~\eqref{eq_output_signal_1}, we then find the corresponding output signal:
\begin{equation}
    I_{Q}({\bf q}_p + {\bf q},\omega) = \gamma_{\rm cav}^2  |G_{13}({\bf q},\omega)|^2 / \pi ~.
\end{equation}
This expression agrees with that derived in \cite{Busch_2014}. The characteristic scattering rate between the condensate photonic fraction and the external photonic vacuum is thus fixed by $\gamma_{\rm cav}$, estimated in our experiment to $\gamma_{\rm cav}\simeq 25\mu$eV.} $G_{13}({\bf q},\omega)$ can be  made explicit using the Bogoliubov transformation Eq.~\eqref{eq_4x4_bogoliubov_transformation} which leads to 
\begin{eqnarray}
& & I_{Q}({\bf q}_p + {\bf q},\omega) = \gamma_{\rm cav}^2  \times \nonumber \\
& & \left[\frac{|u_{c,{\bf q}}v_{c,-{\bf q}}|^2 / \pi}{(\omega - \omega_{{\bf q}_p + {\bf q}})^2 + \gamma_{{\bf q}_p + {\bf q}}^2} + \frac{|u_{c,-{\bf q}}v_{c,{\bf q}}|^2 / \pi}{(\omega + \omega_{{\bf q}_p - {\bf q}})^2 + \gamma_{{\bf q}_p - {\bf q}}^2}\right] ~. \label{eq_I_quantum}
\end{eqnarray}
for the lower polariton resonance, where we have omitted the `$lp$' label. In this expression we have assumed that the normal ($\omega=\omega_{{\bf q}_p + {\bf q}}$) and ghost ($\omega=-\omega_{{\bf q}_p - {\bf q}}$) modes are well split in frequency as compared to $\gamma_{{\bf q}_p\pm {\bf q}}$. 
 \mr{In agreement with \cite{Busch_2014}, we recover the fact that  
 in this regime, the normal branch at ${\bf q}_p + {\bf q}$ and the ghost branch at ${\bf q}_p - {\bf q}$ exhibit an equal emission brightness.} This property is a fundamental consequence of the fact that in this regime, the photons are produced in (quantum-entangled) pairs of ${\bf q}_p \pm {\bf q}$ momenta via Hamiltonian terms equivalent to $\hat a_{{\bf q}_p} \hat a_{{\bf q}_p} \hat a_{{\bf q}_p + {\bf q}}^\dagger \hat a_{{\bf q}_p - {\bf q}}^\dagger$, which destroy two condensate photons (at frequency $\omega_{\rm las}$) to produce a pair of correlated photons in the normal and ghost branches at frequencies $\omega_{\rm las} \pm \omega$.

\subsubsection{\mr{Phonons thermal and vacuum fluctuations} contribution to $I({\bf q},\omega)$}
We now focus on the contribution to the emission intensity originating from coupling to thermal lattice phonons, namely from $\hat f_{\bf q}(\omega) - \hat f_{-\bf q}^\dagger(\omega)$ in Eq.~\eqref{eq_output_signal_2}. We follow a similar derivation as in the \mr{photon vacuum fluctuations} case, using the relation $\hat f_{-\bf q}^\dagger(\omega) = [\hat f_{-\bf q}(-\omega)]^\dagger$. We also use the fact that as $v_s$, the acoustic speed of sound, is much smaller than the speed of light, the phonons involved in the interaction have a wavector $k_z \gg |{\bf q}|$, so that $\omega = c|k_z|$ and we can replace the $g_{xp}({\bf q}, k_z)$ by $g_{xp}(\omega)$. \mr{The scattering rate between the condensate excitonic fraction and the phonon vacuum can be written as}
\mr{
\begin{equation}
    n_x\gamma_{xp}(\omega) = \pi n_x [g_{xp}(\omega)]^2\rho'_{{\bf q},\omega}
    \label{eq_gamma_xp}
\end{equation}
with $\rho'_{{\bf q},\omega} = \sum_{k_z} \delta(\omega - \omega_{{\bf q}, k_z}^{(\rm ph)}) = (A L_z) / (2\pi v_s)$ the reduced phonon density of states,
}
\mr{where $L_z$ is the quantum well thickness and $A$ a quantization area. An explicit expression of $n_x\gamma_{xp}(\omega)$ as a function of the experimental parameters is given in Appendix \ref{app_gamma_xp}. Its maximum value $\gamma_{xp}^M$ provides an order of magnitude of the exciton-phonon interaction strength, and amounts to $\hbar\gamma_{xp}^M\simeq 5\times 10^{-3}\mu$eV$\mu$m$^2$ in our experiment. Assuming a density $n_x\sim 5\times 10^{10}$cm$^{-2}$, it leads to an interaction rate between the condensate excitonic fraction and the phonon vacuum of $\hbar\gamma_{xp}^Mn_x\simeq2.5 \mu$eV, which is at about 10 times smaller than $\hbar\gamma_{\rm cav}$ that determines the scattering rate between the condensate photonic fraction and the photon vacuum. In the generic case where the phonon bath assumes a non-zero temperature, the scattering rate between the condensate excitonic fraction and the phonon thermal bath is enhanced by a thermal phonon population factor, and is given by $n_x\gamma_{xp}(|\omega|)|n_T(\omega)|$, where $n_T(\omega) = 1 / [e^{\hbar \omega / (k_B T)} - 1]$ is the Bose-Einstein distribution. }

\mr{We derive the spectral function of elementary fluctuations generated by thermal phonons at any temperature as} 
\begin{eqnarray}
I_P({\bf q}_p + {\bf q},\omega) &=& \gamma_{\rm cav} \gamma_{xp}(|\omega|) n_x |n_T(\omega)|\times \nonumber \\ 
&&|G_{12}({\bf q},\omega) - G_{14}({\bf q},\omega)|^2 / \pi
\label{IpG}
\end{eqnarray}
As before, we obtain an explicit expression for the matrix elements $G_{12}({\bf q},\omega) - G_{14}({\bf q},\omega)$ using the Bogoliubov transformation [Eq.~\eqref{eq_4x4_bogoliubov_transformation}]. Within the same assumptions as in the \mr{photon vacuum fluctuations} case, we thus obtain for the lower polariton: 
\begin{eqnarray}
I_P({\bf q}_p + {\bf q},\omega) =\gamma_{\rm cav} \gamma_{xp}(|\omega|) n_x |n_T(\omega)| |u_{x,{\bf q}} - v_{x,-{\bf q}}|^2 \times \nonumber\\
\left[\frac{|u_{c,{\bf q}}|^2 / \pi}{(\omega - \omega_{{\bf q}_p + {\bf q}})^2 + \gamma_{{\bf q}_p + {\bf q}}^2} +
\frac{|v_{c,{\bf q}}|^2 / \pi}{(\omega + \omega_{{\bf q}_p - {\bf q}})^2 + \gamma_{{\bf q}_p - {\bf q}}^2} \right] ~.~~~~~~
\label{IpL}
\end{eqnarray}
Note that due to the frequency-dependent factor $\gamma_{xp}(|\omega|)$ as well as to the thermal factor $|n_T(\omega)|$, this lineshape is not purely Lorentzian. This aspect turns out to be significant in developing a careful analysis of the experimental data as discussed below.

Equation (\ref{IpL}) is an important outcome of our analysis. For the normal mode (first Lorentzian), the output photons are produced via the radiative relaxation of an elementary excitation produced itself by the absorption of a thermal phonon at (${\bf q},\omega$) by the condensate. These events occur at a rate which is proportional to $\gamma_{xp}(\omega)n_x$, to the thermal population of lattice phonons $n_T(\omega)$, and to the density of states of the Bogoliubov excitation in the normal mode at $\omega_{\rm las}+\omega$ ($|u_{c,{\bf q}}|^2$ factor). For the ghost branch (second Lorentzian), the output photons are produced by radiative relaxation of an elementary excitation produced itself by spontaneous and stimulated emission of a lattice phonon by the condensate ($|n_T(\omega<0)|=1+n_T(-\omega)$ term), where the density of Bogoliubov excitation in the ghost branch is given by the $|v_{c,{\bf q}}|^2$ factor. Therefore, in this thermal phonons regime, the two Bogoliubov excitation branches have different intensities. \mr{In particular, at $T=0$ the intensity in the ghost branch does not vanish due to the spontaneous emission of lattice phonons in the phonon vacuum, while the intensity vanishes in the normal branch.} This is different from the \mr{regime dominated by photon vacuum fluctuations}, and it will turn out to be a key feature to identify in the experiment the physical origin of Bogoliubov excitations.

As will also be discussed more extensively in section \ref{FB_section}, the overall emission rate is also modulated by a $|u_{x,{\bf q}} - v_{x,-{\bf q}}|^2$ factor, which quantifies the density fluctuation fraction of a Bogoliubov excitation (the other fraction consisting of phase fluctuation). It is this vanishing density fluctuation fraction in Bogoliubov excitations for vanishing momenta that leads to decoupling between the condensate and lattice phonons at small $|{\bf q}|$.

\begin{figure*}[t]
\includegraphics[width=\textwidth]{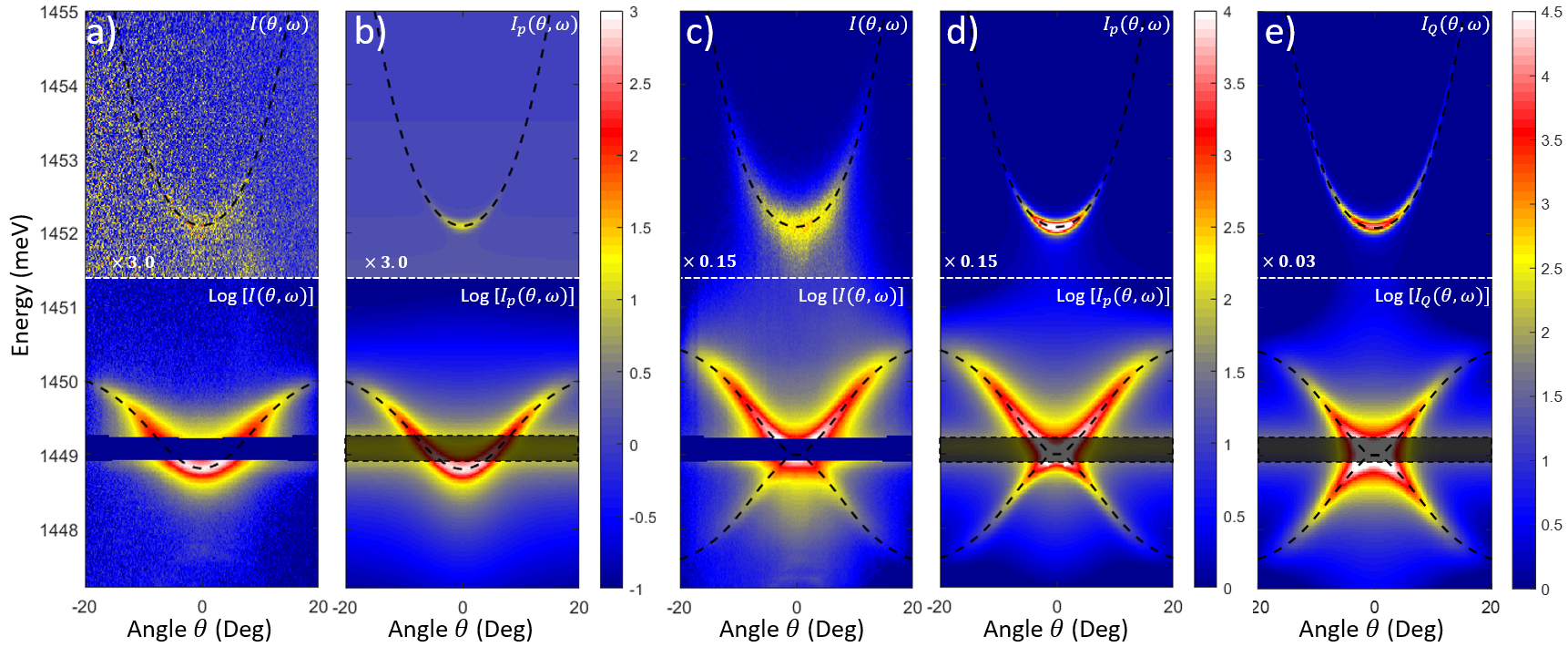}
\caption{ Measured (a and c) and theoretical (b,d,e) spectral functions $I(\theta,\omega)$ of a resonantly-driven polariton condensate in the regime of vanishing interactions (a-b) and in the interacting regime (c,d,e) characterized by $\hbar g_sn_x=0.19\,$meV. The lower polariton spectral area (below the white dashed line) is plotted in Log color-scale, while the upper area is plotted in a linear color scale with the indicated correction factors. The dashed black line is the theoretical dispersion relation of the Bogoliubov excitations in both situations. The missing blue stripe (a and c) and gray rectangles (b, d and e) shows the area rejected from the detection by the spectral filter. In (d) the condensate is assumed to be coupled only to the thermal solid-state phonons with temperature $T_f=15\,$K, as determined from the quantitative analysis of the experiment (see main text). On the contrary, in (e) the elementary excitations are calculated assuming \mr{photon vacuum fluctuations} only.}
\label{f2}
\end{figure*}

\section{Experiment}
\label{section_experiment}
To test these theoretical predictions experimentally, we designed a GaAs-based microcavity in the strong coupling regime containing a single quantum well. This configuration suppresses the dark exciton states that arise in multi quantum well structures \cite{Ivchenko_1994,Savona_1999,Bleu_2020}, and that contribute an unwanted electronic reservoir that strongly perturb the Bogoliubov transformation \cite{Walker_2017,Stepanov_2019}. Another contribution to this reservoir is suppressed by choosing a quantum well thicker than usual ($17\,$nm), which results in a narrower excitonic inhomogeneous broadening, and hence to a lower density of states of localized excitons close to resonance with the lower polariton mode \cite{Savona_2007}.

In the experiment, we excite the planar microcavity, with continuous wave (CW) laser light quasi-resonant with the zero-momentum state of the lower polariton (LP) branch. The upper polariton (UP) branch is blue-shifted above the LP by a Rabi splitting $\hbar\Omega=3.28\,$meV. By taking advantage of the intentional wedge in the cavity thickness, we choose to address a lower polariton (LP) state that has an excitonic fraction of $|X|^2=0.53$. This choice offers significant two-body interactions while preserving a narrow spectral linewidth of the LP states. The microcavity temperature (that of lattice phonons) can be tuned between $T_c=6.6$ and $12\,$K in the vacuum chamber of a Helium flow cryostat. We set the laser frequency at $\omega_{\rm las}=\omega_{lp}({\bf q}_p)+0.20\,$meV$/\hbar$ above our target LP state, where $\omega_{lp}({\bf q})$ is the non-interacting LP dispersion relation and $\hbar\omega_{\rm las}=1449.0\,$meV, and tune its angle close to normal incidence (measured to $\theta_p=0.1\,$degree), which corresponds to a polariton condensate with in-plane momentum $|{\bf q}_p|=0.015\,\mu$m$^{-1}$. For both emission and absorption, the extracavity photon incidence angle is related to the in-plane momentum of the corresponding polaritonic state via $ |{\bf q}|= \omega_{\rm las}\sin(\theta)/c$. Using beam shaping, the laser spot at the microcavity surface (and hence the polariton condensate diameter) amounts to $30\,\mu$m, i.e. much larger than the condensate healing length $\xi=\hbar/\sqrt{2m\Delta_{\rm int}}\simeq 1.5\,\mu$m, where $m=6.6\times10^{-35}\,$kg is the LP effective mass, and $\Delta_{\rm int}\simeq 0.2\,$meV is the two-body interaction energy involved in our experiments. A $20\,\mu$m diameter circular spatial filter selects the central part of the condensate, where the condensate density profile is practically constant.

The driven polariton state has a narrow spectral linewidth $\hbar\gamma_{lp}(\theta\sim 0)=\hbar\gamma_{0}\simeq 0.03\,$meV in the non-interacting regime, as measured after deconvolution of the instrument, such that $\omega_{lp}/\gamma_{0} > \sqrt{3}/2$. In this regime, the driven lower polariton condensate intensity $|\psi_{lp}|^2$, and hence the transmitted light intensity $I_t\propto|\psi_{lp}|^2$, exhibits a hysteretic response as a function of the laser intensity $I_L$ \cite{Baas_2004}. The measured $I_t(I_L)$ is shown in the SI section I \cite{SI}. In the high density state of the bistable region of $I_t(I_L)$, the laser is resonant with the blue shifted LP state and a regime of significant two-body interactions is reached ($\Delta_{\rm int}\geq\hbar\gamma_0$). In the low density state of the bistability, the laser impinges the low energy tail of the essentially unshifted LP state and the two-body interactions is measured to be 400 times smaller (see SI section I \cite{SI}).

\subsection{Measured spectral function of elementary excitation $I(\theta,\omega)$}

\noindent\textit{Non-interacting regime.--} In order to characterize at best the system parameters which do not depend on interactions, we first tune the laser in the regime of vanishing interactions. The emission of the condensate and its elementary excitations, resolved both in angle ($\theta$) and frequency ($\omega/2\pi$), is collected on the transmission side of the microcavity. The condensate emission intensity $I_t$, which is peaked at $\omega=\omega_{\rm las}$ and $k={\bf q}_p$, is several orders of magnitude brighter than the spontaneous emission of Bogoliubov excitations $I(\theta,\omega)$; we reject this signal by exploiting the spectrally narrow character of $I_t$, in using a custom-built image-preserving notch filter of $\sim 0.2\,$meV bandwidth. Filtering out the very bright condensate signal allows us to measure $I(\theta,\omega)$ for both positive and negative frequency with respect to $\omega_{\rm las}$.

The result is shown in Fig.\ref{f2}.a in log (linear) scale for the lower (upper) polariton branch. The emission-free stripe situated between $1448.95\,$meV and $1449.2\,$meV is the result of the above-mentioned filter rejection band. In this non-interacting regime, the elementary excitations have a purely particle (i.e polaritonic) nature: they correspond to the excitation of a polariton out of the condensate into any other polariton state, including the UP. As expected, we do not observe any emission from the ghost mode (at $\omega<\omega_{\rm las}$). Notice that the emission of the UP is much dimmer than that of the LP, as expected in an thermally-assisted excitation mechanism, as the creation of an upper polariton requires the absorption of a phonon of energy $>3.05\,$meV, while at $T_c=6\,$K the phonon thermal distribution falls off exponentially with the decay constant $k_BT_c=0.56\,$meV. We further elaborate below on the role of thermal phonons in the creation of elementary excitations.  \\

\noindent\textit{Interacting regime.--} We then increase the laser intensity in order to reach the high-density state of the bistability, where the interactions are significant. In this regime, the interaction energy amounts to $\Delta_{\rm int}=0.19\,$meV. The measured $I(\theta,\omega)$ is shown in Fig.\ref{f2}.c. As is immediately apparent, with respect to the non-interacting case (panel a of the figure), the interactions modify significantly the elementary excitations dispersion relation. Two different branches can be identified, the one with a positive effective mass and frequency range above $\omega_{\rm las}$ is the Bogoliubov excitations normal (N) branch $\hbar\omega_{N}(\theta)$. The dimmer one, characterized by a negative effective mass, and frequency range below $\omega_{\rm las}$ is the ghost (G) branch $\hbar\omega_{G}(\theta)$. The shape of $\hbar\omega_{N,G}(\theta)$ is also clearly modified (i.e. steeper) with respect to the free particle dispersion relation $\hbar\omega_{lp}(\theta)$. These features fully agree with the expected characteristics of the dispersion relation for LP Bogoliubov excitations. Finally, the UP dispersion relation is also clearly visible, and its shape is left essentially unchanged with respect to the non-interacting regime, as we will see more quantitatively later on, and which is expected in the regime where $\Omega\gg \Delta_{\rm int}$ (See Appendix section \ref{appendix_u_and_v}).

\mr{The emission from a ghost branch has been reported once under the intrinsic fluctuations of the system but for a much lower interaction strength \cite{Zajac_2015}. It has been also observed in pump-probe experiments \cite{Kohnle_2011,Claude_2022}. A ghost branch of elementary excitations can also emerge in situations in which the condensate coherence results not from the laser laser coherence, but from spontaneous symmetry breaking \cite{IC_2013}. Such condensates have specific fluctuations \cite{Fontaine_2022}, and their elementary excitations involve a different Bogoliubov-like transformation, characterized by a diffusive Goldstone mode at vanishing frequency \cite{Wouters_2007,Ballarini_2020}, as well as a ghost mode \cite{Pieczarka_2020,Pieczarka_2021,Steger_2021}.}\\

\noindent\textit{Data analysis.--} In order to extract the different relevant observables from these measured $I(\theta,\omega)$, such as in particular the dispersion relation of the different modes $\omega_j(\theta)$, ($j=\{{\rm lp,up}\}$ in the non-interacting regime and $j=\{{\rm N,G,up*}\}$ in the interacting regime, where ${\rm up}^*$ labels the normal branch of the upper polariton), as well as the peak-integrated intensity $A_j(\theta)$ of those modes, we first use the measurement in the non-interacting regime carried out at different temperatures (between $T_c=6.6\,$K and $T_c=12\,$K). From these data, we can precisely determine most experimental parameters such as e.g. the bare exciton and cavity energies, the Rabi splitting, as well as the cavity photon effective mass. Moving then to the interacting regime, we fit the experimental $I(\theta,\omega)$ using Eq.~(\ref{IpG}), keeping fixed these parameters that do not depend on interactions, and introduce three additional $\theta$-dependent fit parameters, which allow us to freely capture both the energy and the amplitude of each individual peak in the experiment. As further discussed in the Supplementary Information \cite{SI}, this method provides us with the experimental values of $\omega_j(\theta)$, and $A_j(\theta)$ that can then be compared with the theory, allowing us to estimate the remaining interaction-dependent parameters of the model.\\

\noindent\textit{Theoretical spectral functions.--} In the non-interacting regime, $|v_c|^2=0$; from Eq.~(\ref{eq_I_quantum}) it 
follows that the \mr{photon vacuum fluctuations} cannot create any elementary excitation, \mr{while thermal phonons can}. The corresponding theoretical spectral function $I(\theta,\omega)$ is shown in Fig.\ref{f2}.b and is in excellent agreement with the measured spectral function (panel a of the figure). In the next section, we will proceed to a quantitative comparison between theory and experiments. We next move to the interacting regime and keep the same model parameters as in the non-interacting case, except for a nonzero value of the nonlinearity $\hbar g_sn_x=0.19\,$meV, and a higher temperature $T=15\,$K$>T_c$. \mr{This temperature higher than that of the cryostat, is likely the result of residual absorption of the high intracavity field by the GaAs alloys, which produces a heating power due to ensuing non-radiative recombination.} By repeating the same analysis as above, namely assuming only contribution from thermal lattice phonons yields the theoretical spectral function shown in Fig.\ref{f2}.d. Also in this case we find  an extremely good agreement with the experimental spectral function (panel c of the figure). The hypothesis of neglecting the \mr{photon vacuum fluctuations} will be justified in detail in Sec.~\ref{sec_Atheta} by the analysis of the spectral intensities.\\

\noindent\textit{Thermal fluctuations versus \mr{photon vacuum fluctuations}.--} As discussed in Sec.~\ref{section_theory}, the generation of elementary excitations by thermal phonons or by \mr{photon vacuum fluctuations} result in a very different $I(\theta,\omega)$ pattern: Fig.~\ref{f2}.(b,d) show the calculated spectral functions including only excitations via thermal phonons, while  Fig.\ref{f2}.e shows the corresponding pattern calculated assuming only quantum photon fluctuations. The reason for this difference is that in the quantum regime, excitations are only created in correlated pairs, implying that  -- for ${\bf q}_p=0$ -- $I(\theta,\omega)$ and $I(-\theta,-\omega)$ have an equal brightness.
In the regime dominated by thermal lattice phonons, the mechanism generating elementary excitations is not expected to produce as much pair correlations. We notice however that a finite amount of correlations is still expected in this regime, as the lattice phonon which is emitted for the creation of an excitation at $(-\omega,-\theta)$ has precisely the right energy and momentum for the creation, via its absorption, of a second excitation at $(\omega,\theta)$. A quantitative account for the correlation properties of the spontaneous emission signal, both on the theory and experimental level, exceeds the scope of the present study, but represents a very clear future research direction. In the next sections, we proceed to a quantitative analysis of the experimental data.

\subsection{Dispersion relations $\omega_j(\theta)$}
\begin{figure}[bt]
\includegraphics[width=\columnwidth]{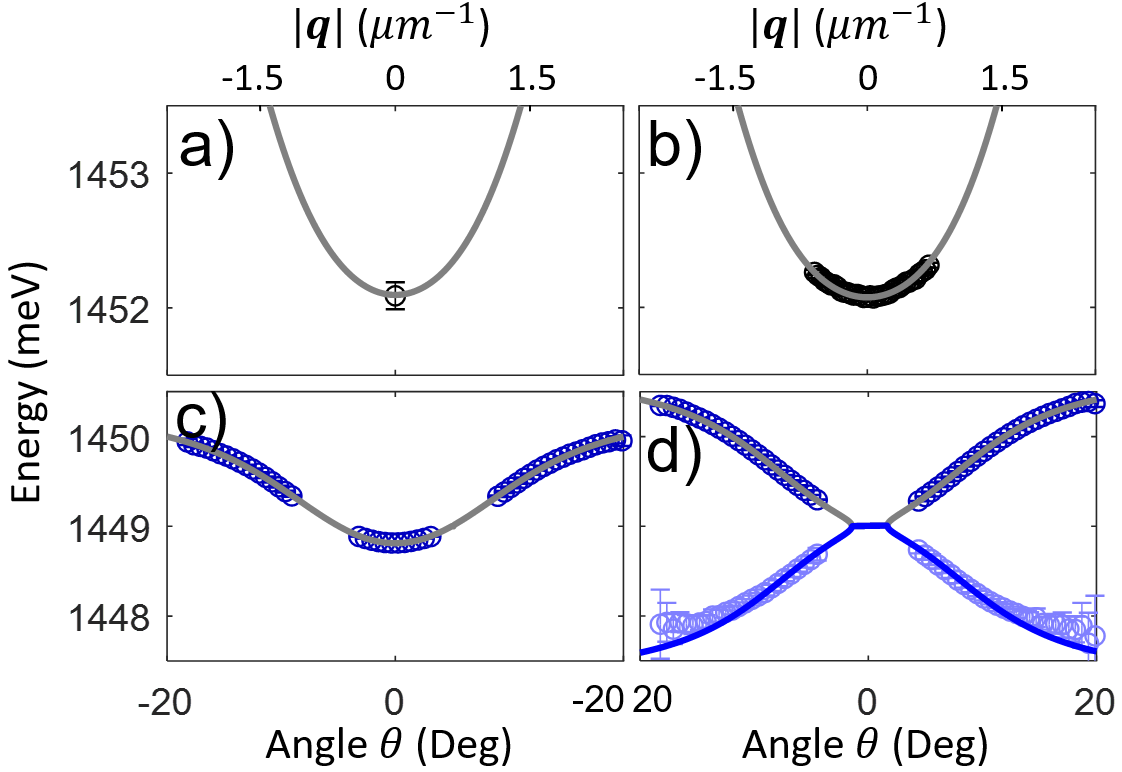}
\caption{Dispersion relations in the non-interacting (a,c) and interacting (b,d) regime at $T_c=6.6K$. The data points extracted from the analysis of the measured spectral functions $I(\theta,\omega)$ are shown in black for the upper polariton branch, in dark (light) blue for the normal (ghost) mode of the lower polariton, with error bars determined from the fitting procedure and corresponding to the $1\sigma$ confidence interval. The solid lines show the theory.}
\label{f3}
\end{figure}

\noindent\textit{Comparison between experiment and theory.--} From the fitting procedure of the spectral function $I(\theta,\omega)$ as described in the previous section, we extract the experimental dispersion relations of the elementary excitations. In the non-interacting regime, the result is shown in Fig.\ref{f3}.a and Fig.\ref{f3}.c for both the upper [$\omega_{up}(\theta)$] and lower [$\omega_{lp}(\theta)$] polariton modes. In the interacting regime, three different modes are extracted: $\omega_G(\theta)$, $\omega_N(\theta)$ (Fig.\ref{f3}.d) for, respectively, the ghost and normal modes of the lower polariton and $\omega_{up*}(\theta)$ (Fig.\ref{f3}.b) which corresponds to the upper polariton branch that retains its free particle character in the limit $\hbar\Omega\gg\Delta_{\rm int}$ \cite{Ciuti_2005}. The theoretical dispersion relations obtained from diagonalizing the $M_{\bf q}$ matrix Eq.~(\ref{eq_M_mat}) are shown as the solid lines, using the parameters of the 'hotter' regime described above, and by adjusting the two coefficients $g_s$ and $g_x$ that describe the interactions in the polariton gas.

We obtain an excellent quantitative agreement between theory and experiment, which allows us to draw several important conclusions. Firstly, the fact that the theory accurately reproduces both (i) the N and G branches spectral blueshift, and (ii) the slope of $\omega_{N,G}(\theta)$ at low momenta, implies that no additional reservoir is perturbing the Bogoliubov transformation. Indeed, we have shown both experimentally and theoretically in previous works \cite{Stepanov_2019,Amelio_2020}, that a reservoir of particles much heavier than polaritons (such as e.g. electron-hole pairs, or dark excitons) coexisting with the condensate would induce an additional blueshift $\Delta_{\rm res}$ to the dispersion relation, but would not modify the speed of sound $c_s$, for which the relation $c_s=\sqrt{\Delta_{\rm int}/m}$ remains valid. Since the experimentally observed blueshift is $\Delta_o=\Delta_{\rm int}+\Delta_{\rm res}$, in presence of a reservoir, one finds that $c_s<\sqrt{\Delta_o/m}$. Note that this analysis is also valid for non-sonic states, in which the dispersion relation slope is also determined only by $\Delta_{\rm int}$, and thus also leads to a similar inequality when $\Delta_o>\Delta_{\rm int}$, something that can be verified numerically. This analysis thus shows that within the experimental uncertainty, the measured elementary excitations involve no such reservoir. 


\mr{We note that the best fitted dispersion relation shows a flat segment around $\theta = 0$, which typically indicates the proximity of a dynamical instability. This is consistent with the experiment as, while we do not observe any signature of instability, we indeed operate the system close to an instability point, namely the switch down point of the upper branch of the hysteresis curve (the working point is labelled '3' in Fig.1 of the supplemental material). This working point provides indeed the best compromise between high interaction energy and low laser power. In practice, the onset of instability is determined by the positivity of the largest eigenvalues imaginary part $\Gamma_{B,m}$ \cite{IC_2004}. Our fit yields $\Gamma_{B,m}=+50^{+85}_{-65}\,\mu$eV, which is compatible with stable solutions within the experimental errorbars.}

\noindent\textit{Excitonic nonlinearity.--}
The best fit obtained above for the dispersion relation in the interacting case involves the ratio  $g_xn_x/g_sn_x$ of the nonlinear contributions, where $n_x$ is the excitonic fraction of the condensate density. The fact that all three dispersion relation branches (G,N,up*) are experimentally accessible provides us with a unique mean to determine this ratio unambiguously. Indeed, while both $g_x$ and $g_s$ result in a blueshift of the lower polariton N and G modes, $g_x$ would also blueshift the up* branch with respect to its non-interacting counterpart; on the other hand $g_s$ leaves this up* branch spectrally unshifted, and leads to an effectively reduced Rabi splitting. This qualitative difference allows us to clearly separate both $g_s$ and $g_x$ contributions, and yields the ratio $g_x/g_s\simeq 0^{+0.3}_{-0}$ where $0$ is the best fit, and $\hbar g_sn_x=0.19\,$meV. This result is of significant interest, and its microscopic interpretation deserves a detailed analysis that exceed the scope of this work.

Let us mention that the fact that $g_s$  seems to dominate over $g_x$ does not contradict previous works dealing with polaritonic nonlinearities, since as explained above, $g_s$ and $g_x$ contribute to blueshifting the LP mode which is the only focus of nearly all studies on exciton-polaritons, while the upper polariton mode is typically disregarded. 

\subsection{Emission intensity of elementary excitations}
\label{sec_Atheta}
\begin{figure}[t]
\includegraphics[width=\columnwidth]{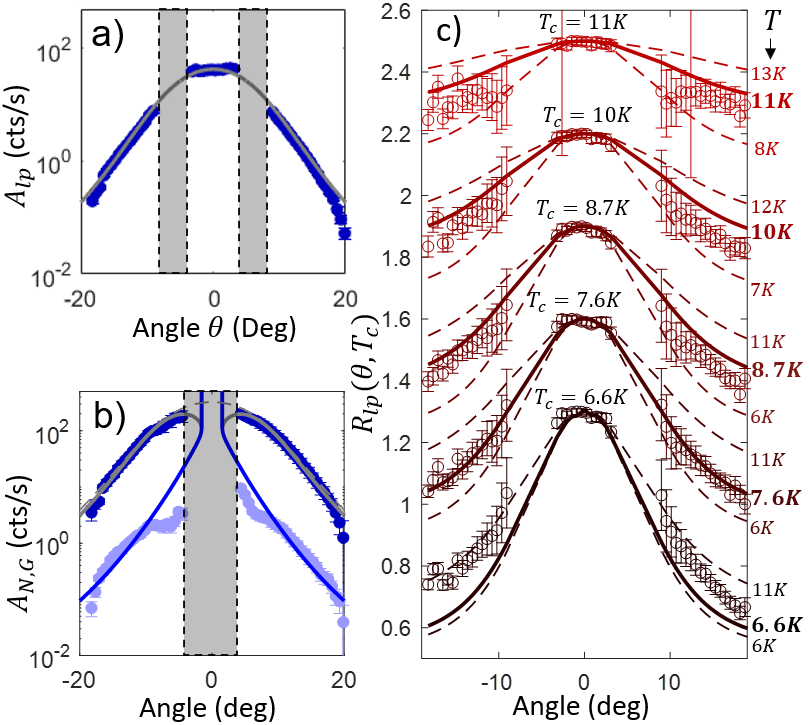}
\caption{Analysis of integrated emission intensity $A_j(\theta)$. Measured and theoretical $A_j(\theta)$ at $T_c=6.6K$ (a) in the non-interacting regime ($j=$lp) and (b) in the interacting regime, where the normal (ghost) mode $j=$N ($j=$G) corresponds to the upper (lower) curve of dark (light symbols). The theory is shown as solid lines. The gray rectangle indicates the area of inaccessible measurements due to the spectral filter rejection in the experiment. (c) Measured integrated intensity ratio $R_{lp}(\theta)[T_c]=A_{lp}(\theta)[T_c]/A_{lp}(\theta)[T_c=12K]$ for $T_c=[7.6,8.7,10,11]\,$K (symbols) in the non-interacting regime. The theory assuming the nominal phonon temperature $T=T_c$ is shown as a solid line, as well as two more temperatures: a lower and a higher with $5\,$K difference, which are shown as dotted lines }
\label{f4a}
\end{figure}
We then extract $A_j(\theta)$ from the measured spectral function $I(\theta,\omega)$, which is defined as the spectral integration of the emission intensity over the peaks of each mode $j=$N,G, as labelled in Eq.~(\ref{IpL}). $A_j(\theta)$ is the right observable to identify the origin of the fluctuations generating the elementary excitations (namely, thermal lattice phonons or \mr{photon vacuum fluctuations}). Assuming narrow Lorentzian lineshapes, Eq.~(\ref{IpL}) yields a simple approximate expressions for $A_j(\theta)$ in the regime dominated by thermal lattice phonons, that reads:
\begin{eqnarray}
A_{N}(\theta) & \simeq & \gamma_{\rm cav} \gamma_{xp}[\omega_N(\theta)] n_x n_T[\omega_N(\theta)]|u_c(\theta)|^2 \nonumber \\
& \times & |u_{x}(\theta) - v_{x}(\theta)|^2/\gamma_N(\theta), 
\label{ANp}
\end{eqnarray}
and, similarly:
\begin{eqnarray}
A_{G}(\theta) & \simeq & \gamma_{\rm cav} \gamma_{xp}[-\omega_G(\theta)] n_x (1+n_T[-\omega_G(\theta)])|v_c(\theta)|^2 \nonumber \\
& \times & |u_{x}(\theta) - v_{x}(\theta)|^2/\gamma_G(\theta), 
\label{AGp}
\end{eqnarray}
where $-\omega_G(\theta)>0$ and we assumed that ${\bf q}_p\approx 0$. In the \mr{regime dominated by photon vacuum fluctuations}, Eq.\eqref{eq_I_quantum} yields:
\begin{equation}
A_N(\theta)=A_G(\theta)\simeq\gamma_{\rm cav}^2 |u_{c}(\theta)v_{c}(\theta)|^2/\gamma_N(\theta).
\label{AQ}
\end{equation} 
In our dataset, the N mode is found much brighter than the G mode, which is in agreement with Eqs.~(\ref{ANp},\ref{AGp}), and not with Eq.~\eqref{AQ}. This feature rules out a dominant contribution of \mr{photon vacuum fluctuations} in this experiment.

The quantitative analysis of $A_j(\theta)$ is shown for a phonon temperature $T_c=6.6\,$K, both in the non-interacting case [Fig.~\ref{f4a}.a] and in the interacting case [Fig.~\ref{f4a}.b]. In the non-interacting case, we observe a decay of $A_{lp}(\theta)$ as a function of $\theta$ which is in very good agreement with the theoretical prediction. The latter also agrees quantitatively with the theory in the interacting case assuming $T=15\,$K as explained above. Importantly, the relative intensity between $A_G(\theta)$ and $A_N(\theta)$ is captured strikingly well in this log-scaled plot, except at small angles where $A_G(\theta)$ is found to deviate from the theory. Indeed at smaller angles, the Bogoliubov excitation peaks become increasingly harder to separate from an additional emission contribution that we attribute to spatial inhomogeneities. This agreement shows that the generation of elementary excitations is most likely dominated by thermal phonon in our experimental conditions.

We confirm this important feature by analyzing the ratio $R_{lp}(\theta)[T_c]=A_{lp}(\theta)[T_c]/A_{lp}(\theta)[T_{0}]$ in the non-interacting case, where $T_0$ is a reference temperature, and $A_{lp}(\theta)[T_c]$ is the measured lower polariton intensity at another cryostat temperature $T_c$. According to Eq.~(\ref{ANp}), $R_{lp}(\theta)$ is free from any temperature-independent parameters such as the non-trivial energy-dependent exciton-phonon interaction $g_{xp}(\omega)$, that involves the excitonic envelope wavefunction, and it depends only on temperature as $R_{lp}(\theta)[T_c]=n_{T_c}(\omega)/n_{T_0}(\omega)$ where $n_{T_c}(\omega)$ is the Bose-Einstein distribution of temperature $T_c$ evaluated at $\omega=\omega_{lp}(\theta)$. In the experiment, $R_{lp}(\theta)[T_c]$ is not as simple, as the excitonic transition energy, and to a lesser extent the cavity mode, both redshift for increasing temperature. This redshift modifies the LP dispersion relation $\omega_{lp}(\theta)$ and the excitonic fraction. However $R_{lp}(\theta)$ is still an observable that involves a lower number of complex experimental parameters. We thus perform this analysis varying the cryostat temperature in the range $T_c=[6.6,7.6,8.7,10,11,12]\,$K, and choose $T_c=12$ as the reference temperature $T_0$. The result is shown in Fig.\ref{f4a}.c, and compared with the theoretical prediction for $R_{lp}(\theta)[T_c]$ including the temperature-dependent parameters mentioned above. We analyzed the experimental uncertainty (see the dashed lines in Fig.\ref{f4a}.c) and found that the temperature estimate is accurate within $5\,$K. Since this accuracy is comparable with the investigated temperature range, a fully quantitative comparison is not possible, however the observed trend is clearly consistent with a regime dominated by thermal phonons. The two analyses presented in this section thus demonstrates consistently that in our experiment, the Bogoliubov excitations result dominantly from the interaction of the condensate with thermal lattice phonons.

\subsection{Characterization of Bogoliubov amplitudes}
\label{section_bogoliubov_amplitudes}

\begin{figure}[t]
\includegraphics[width=\columnwidth]{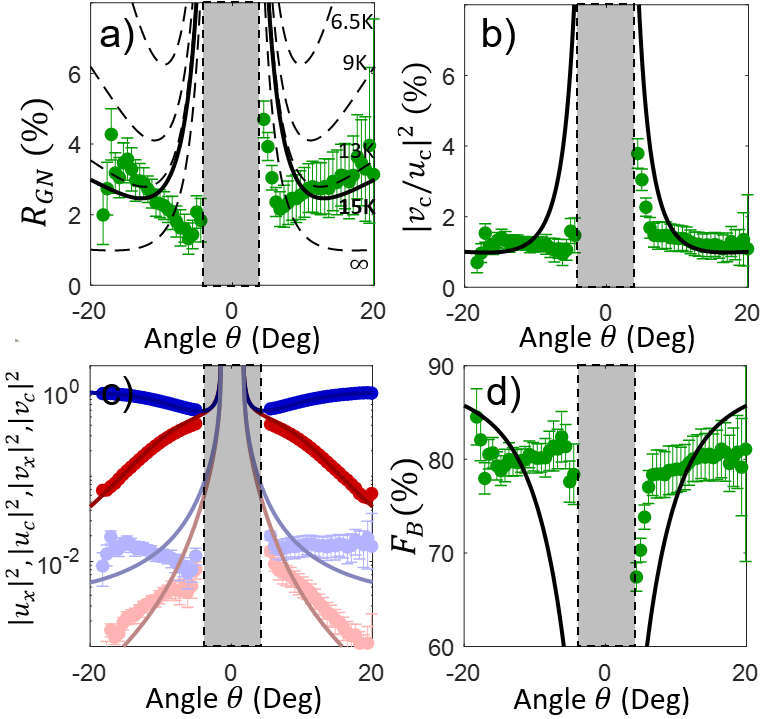}
\caption{Bogoliubov amplitudes analysis. (a) $R_{GN}(\theta)$, (b) photonic Bogoliubov squared amplitude ratio $|v_c(\theta)/u_c(\theta)|^2$. The four squared Bogoliubov amplitudes are plotted in (c) in semilog scale with $u_c$ ($v_c$) in dark (light) blue, and $u_X$ ($v_X$) in dark (light) red. (d) Measured and theoretical Bogoliubov correction factor $F_B(\theta)$ to the polariton-phonon interaction rate. The data points in (a,b,d) are shown in green. In all panels, the solid lines show the theory. The dashed line in (b) show the theoretical $R_{GN}(\theta)$ assuming the different phononic temperatures indicated, in addition to the best fit $T=15\,$K which is shown as a solid line. The gray rectangle indicates the area of inaccessible measurements due to the spectral filter rejection in the experiment.}
\label{f4b}
\end{figure}

\noindent\textit{Intensity ratio and Bogoliubov coefficients.--} In our experiment, the fact that the Bogoliubov excitations are created dominantly by interaction with the thermal bath of lattice phonons opens up a unique opportunity to obtain a quantitative estimate of the Bogoliubov coefficients $(u_c(\theta),u_x(\theta),v_c(\theta),v_x(\theta))$, by comparing the emission intensity of the ghost branch and of the normal branch. Indeed, assuming that $\gamma_G(\theta)\simeq\gamma_N(\theta)$, and using the fact that ${\bf q}_p\approx 0$, the intensity ratio $R_{GN}(\theta)=A_{G}(\theta)/A_{N}(\theta)$ assumes the strikingly simple expression 
\begin{equation}
R_{GN}(\theta)=\frac{|v_{c}(\theta)|^2}{|u_{c}(\theta)|^2}  e^{\hbar\omega_N(\theta)/k_BT}\,.
\end{equation}
Experimentally, $A_G$ and $A_N$ are measured in the exact same experimental conditions, as they are obtained in a single shot of the CCD camera, so that no extra $\theta$-dependent factor remains in the ratio $R_{GN}$.

The comparison between the experiment carried out at $T_c=6.6$K, and the corresponding theory, are shown in Fig.\ref{f4b}.a. Considering the experimental uncertainty quantified by the error bars (notice also that the full range of the plot is only $R_{GN}=[0,8]\%$), we obtain a fairly good agreement, regarding both the value of the ratio itself, and its trend as a function of $\theta$. 

In order to illustrate the crucial role of temperature on the brightness of the ghost branch with respect to the normal one, theoretical plots of $R_{GN}(\theta)$ are shown in Fig.\ref{f4b}.a for several other temperatures between $6.5\,$K and $\infty$ (dashed lines). When the temperature is much higher than the highest frequency of the lower polariton mode 
(namely when $k_BT\gg k_BT_{GN}=\hbar \omega_{N}(\theta_{\rm max})$, which in our case corresponds to $T_{GN}\simeq 15\,$K), we have $R_{GN}\simeq|v_c/u_c|^2$ (dashed line labelled as $T=\infty$ in Fig.\ref{f4b}.a). Upon cooling down below $T_{GN}$, less thermal phonons are available to create excitations into the normal branch (starting with the highest energy states, namely largest angles or momenta), so that the ghost branch, which is populated by both spontaneous and stimulated emission of lattice phonons, increases in intensity with respect to the normal branch.

Remarkably, one predicts that for low enough temperature, and neglecting the \mr{photon vacuum fluctuations} to get a simple estimate, the ghost branch emission intensity can even overcome that of the normal branch. For a given $\theta$, this happens when $R_{GN}(\theta) \ge 1$, namely when $k_BT \le \hbar\omega_N / \log(|u_c/v_c|^2)$. For instance, using our experimental parameters and a typical angle $\theta=10\,$degree, we find  that the temperature  for which equal brightness of the two branches is achieved $T_{EB} \lesssim 5.2\,$K, which is about three times colder than the actual phonon temperature estimated for this experiment. This criterion provides a likely explanation regarding the notoriously difficult observation of the ghost mode spontaneous emission in exciton-polariton systems, in addition to the requirement of eliminating any detrimental electronic reservoir.

Finally, it is straightforward to estimate the Bogoliubov coefficient ratio $|v_c/u_c|^2=R_{GN}\times \exp[-E_N/k_BT]$ from $R_{GN}(\theta)$, using the temperature $T=15\,$K. The result is shown in Fig.\ref{f4b}.b together with the theory (solid line). In full agreement with the theory, we find that the photonic fraction of the "hole" character of Bogoliubov excitations ($|v_c|^2$) amounts to $\sim 1.5\%$ at high angle/energy, and up to $\sim 4\%$ for the lowest angles. While this correlation is modest, it is clearly established in our experiment, and the detailed understanding provided by our theoretical model is a guide towards realizing more strongly correlated quantum fluids of light, and in particular generating quantum-correlated pairs of excitations by lowering the temperature to enter a regime dominated by \mr{photon vacuum fluctuations} (see Section \ref{section_discussion}.A).\\

\noindent\textit{Extracting all Bogoliubov coefficients.--} The Bogoliubov transformation is defined by $4$ complex coefficients $(u_c,u_x,v_c,v_x)$, that characterizes the exciton and photon particle fractions (first two), and the exciton and photon hole fractions (last two), of a Bogoliubov excitation (see  section \ref{section_theory}.B and Eq.~\eqref{eq_4x4_bogoliubov_transformation}). In our experimental conditions, these four amplitudes can be determined to within a good approximation from the knowledge of the ratio $|v_c/u_c|^2$ determined above, and from the knowledge of the excitonic $X$ and photonic $C$ Hopfield coefficients of the (non-interacting) polariton states, (their $\textbf{q}$-dependence has been dropped like for $(u,v)$). Indeed as mentioned earlier, the fact that the Rabi splitting is much larger than the interaction energy (namely $\Omega \gg \Delta_{\rm int}$), implies that the lower and upper polaritons do not hybridize under the Bogoliubov transformation, and that the hole fraction of the upper polaritons is vanishing \cite{Ciuti_2005}. One can thus define two lower-polariton-only Bogoliubov coefficients $(u,v)$ that are related to the 4 original coefficient via the relations $(u_x,v_x)/X=(u_c,v_c)/C=(u,v)$, which leads also to $|v_c/u_c|^2=|v/u|^2$ (see Appendix \ref{appendix_u_and_v} for a detailed derivation). In this regime, the Bogoliubov coefficients are approximately real, and using the normalization condition $|u|^2 - |v|^2 = 1$, one can thus derive both $u$ and $v$ from the knowledge of $|v_c/u_c|^2$, and hence all four coefficients $(u_c,u_x,v_c,v_x)$. By doing so, one obtains all the Bogoliubov coefficients plotted on Fig.\ref{f4b}.d, together with their exact theoretical prediction showing a remarkable agreement that further supports the approximations discussed above.

\section{Discussion}
\label{section_discussion}

\begin{figure}[bt]
\includegraphics[width=\columnwidth]{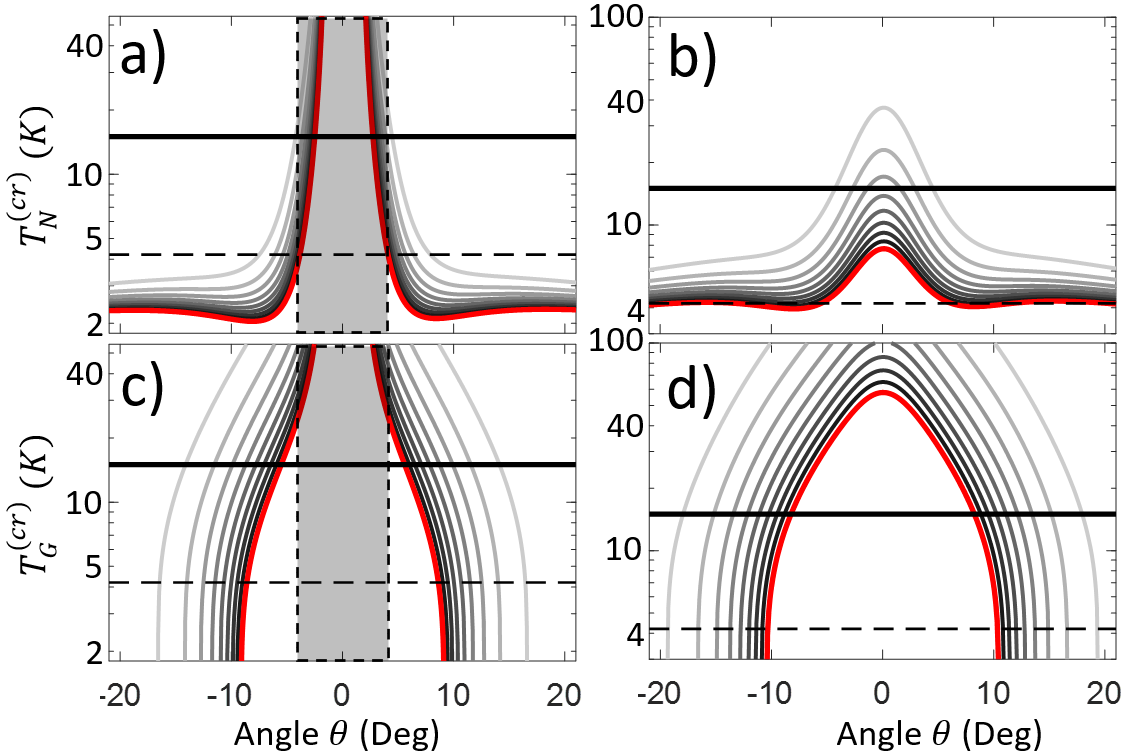}
\caption{Calculated angle-dependent thermal to quantum regime crossover temperature $T_{cr,N}$ for the normal (a) and $T_{cr,G}$ for the ghost modes (c) in semilog scale. The excitonic density is increasing from $n_x=0.1n_{x,M}$ to $n_x=0.9n_{x,M}$ from the lightest to the darkest gray lines. \mr{The thick red line shows the result assuming $n_x=n_{x,M}$, the Mott density, which constitutes a solid upper bound of $n_x$}. The thick horizontal line shows a photon temperature of $T_c=15\,$K, and the dashed horizontal shows liquid Helium temperature $T_{He}=4.2\,$K. panels (a and c) are calculated for the total bluesgift achieved in this work $\hbar g_sn_x=0.19\,$meV, while panels (b and d) are calculated for a twice larger blueshift $\hbar g_sn_x=0.38\,$meV.}
\label{f5}
\end{figure}

\subsection{\mr{Crossover temperature into the regime dominated by photon-vacuum fluctuations}}

\mr{Owing to the thermal population factor entering the interaction between the condensate and the lattice phonons ($n_T(\omega)$ in Eq.(\ref{IpL})), the relative contribution of photon vacuum fluctuations and of the lattice phonons can be tuned to a large extent by changing the lattice temperature $T$. Within the right parameters range, a non-zero crossover temperature can thus be determined, below which the Bogoliubov excitations emission results dominantly from the interaction between the condensate and the photon vacuum fluctuations. In such a regime, the condensate starts to effectively decouple from the lattice, which is most likely a necessary condition for the elementary excitations to generate quantum correlations \cite{Busch_2014,Koghee_2014,Gerace_2012,Nguyen_2015,Steinhauer_2016,Unruh_1981,Jacquet_2020}. A quantitative derivation of their emergence would require calculating two-photon correlations \cite{Busch_2014}, which exceeds the scope of this work.}

\mr{As we shall see in this section, a single crossover temperature cannot be defined for the whole condensate. The highest crossover temperature occurs for low-momenta excitations, and gets lower for higher momenta. Upon cooling down, the condensate decoupling from the lattice is thus a smooth process, in which an increasing range of momenta (or equivalently of emission angle $\theta$) of Bogoliubov excitations become more coupled to the photon vacuum fluctuations than to the lattice.}

The crossover temperature can be simply estimated for the normal branch by equating Eqs.~(\ref{ANp},\ref{AQ}), yielding:
\begin{equation}
T^{(cr)}_{N}(\theta)=\frac{\hbar\omega_N(\theta)/k_B}{\log{\left[1+\frac{\gamma_{xp}(\theta)n_x}{\gamma_{\rm cav}}\frac{|u_x(\theta)-v_x(\theta)|^2}{|v_c(\theta)|^2}\right]}}~.
\label{Tcross_N}
\end{equation}
Similarly, for the ghost mode we equate Eqs.~(\ref{AGp},\ref{AQ}) to obtain:
\begin{equation}
T^{(cr)}_{G}(\theta)=-\frac{\hbar\omega_N(\theta)/k_B}{\log{\left[1-\frac{\gamma_{xp}(\theta)n_x}{\gamma_{\rm cav}}\frac{|u_x(\theta)-v_x(\theta)|^2}{|u_c(\theta)|^2}\right]}} ~.
\label{Tcross_G}
\end{equation}

One difficulty in deriving a quantitative estimate of $T^{(cr)}_{N,G}(\theta)$ is that it requires knowing the excitonic density $n_x$ involved in the condensate, which is not easy to determine in the experiment. However, note that $T^{(cr)}_{N,G}(\theta)$ depends only logarithmically (and hence weakly) on the product $\gamma_{xp}(\theta)n_x$. Regarding $n_x$, we use the fact that it it is necessarily lower than the excitonic Mott density $n_{x,M}$, for which the Rabi splitting, and hence the polaritonic state would be fully collapsed. $n_{x,M}$ depends only on the excitonic Bohr radius as $n_{x,M}\sim 1/(\pi a_B^2)$ and is thus easy to estimate; it amounts to $\simeq 3.2\times 10^{11}\,$cm$^{-2}$ in our quantum well. Concerning  $\gamma_{xp}(\theta)$, a realistic estimate is derived in Appendix \ref{app_gamma_xp}.

The resulting crossover temperature for the normal branch $T^{(cr)}_{N}(\theta)$ is thus plotted in Fig.\ref{f5}.a, by considering a range of excitonic densities between $5\%$ (lightest gray line) and $100\%$ (red line) of $n_{x,M}$. The parameters used in the calculation are those derived for our experiment at $T_c=6.6\,K$, and the total blueshift $\hbar g_sn_x=0.19\,$meV is kept constant (and hence the Bogolibuov coefficients as well) to its measured value. This analysis shows that the normal branch crossover temperature is situated realistically within the interval $[2.5,4.5]$K within a typical \mr{emission angular aperture of $10\,$degrees. }

The ghost branch crossover temperature $T^{(cr)}_{G}(\theta)$ is shown in Fig.\ref{f5}.c. As an obvious difference with the normal branch, no crossover temperature can be determined above a critical angle ($|\theta_{cr}|\sim 10\,$degree in our simulation of this experiment), because above $|\theta_{cr}|$, the \mr{photon vacuum fluctuations} can not overcome the vacuum fluctuations of the lattice phonons that are present at $T=0K$, for the generation of Bogoliubov excitations. \mr{This feature is particularly striking as it means that in this realistic case, the condensate can never be considered decoupled from its solid-state lattice environment \mr{for all possible elementary excitations momenta}. Physically, the generation of Bogoliubov excitations by the lattice at $T=0K$ occurs} into the ghost mode by spontaneous emission of phonons into the phonon vacuum. The corresponding $T=0\,$K ghost branch emission intensity $A_{G,0}$ can be derived from Eq.~\eqref{AGp} as:
\begin{eqnarray}
A_{G,0}& = & A_{G}(\theta,T=0) = \gamma_{\rm cav}\gamma_{xp}(\theta)n_x \nonumber \\
& \times &|v_c(\theta)|^2|u_x(\theta)-v_x(\theta)|^2.   
\end{eqnarray}
When $A_{G,0}$ is larger than that the contribution generated by \mr{photon vacuum fluctuations} [Eq.~\eqref{AQ}], the latter can then never become dominant for the ghost branch. The parameters involved in $A_{G,0}$ show that $|\theta_{cr}|$ can be increased via different parameters such as the total interaction energy, or the excitonic $|X|^2$ and photonic $|C|^2$ fractions (embedded within the Bogoliubov coefficients $(u_x,v_x,u_c,v_c)$) that allow modulating to a large extent the relative coupling of polaritons to phonons on one hand ($\propto |X|^2$) and to free space photons ($\propto |C|^2$) on the other hand. 

\mr{In Fig.\ref{f5}.b-d, we computed the crossover temperature $T^{(cr)}_{N}(\theta)$ and $T^{(cr)}_{G}(\theta)$ for the same range of excitonic densities but considering a twice larger total blueshift $\hbar g_sn_x=0.38\,$meV. This change amounts to increasing the laser intensity substantially and hence changing the working point on the hysteresis curve (Fig.1 of the SI) to the right, where the elementary excitations exhibit a gaped dispersion relation.The benefit of this increase of interaction energy is twofold. The $|v_{x,c}|^2$ coefficient magnitude increase, and the spectral gap in the excitation dispersion relation is of comparable size as the spectral filter in the experiment. As a result, the emission even from the smallest angles, where $|v_{x,c}|^2$ is the largest, becomes experimentally accessible. Note that in practice, increasing the excitonic fraction $|X|^2$ or the laser intensity typically require specific optimization, as it results in spectral broadening of the polariton state and/or in an increased heating due to residual absorption.}

\subsection{Decoupling of low-momenta Bogoliubov excitations from lattice phonons}
\label{FB_section}

One last striking property that we examine in this section, is the overall reduction of the coupling strength between polaritons and the thermal bath of lattice phonons that results from the Bogoliubov transformation. This reduction can be seen already at the Hamiltonian level in which rewriting the exciton-phonon interaction term [Eq.~\eqref{eq_Vxp_bogo}] in the Bogoliubov basis results in a correction of the interaction energy amplitude $g_{xp}(\omega)$ by a factor $(u_{x,{\bf q}}-v_{x,-{\bf q}})$. Assuming a LP-only condensate, which as discussed both in Section \ref{section_bogoliubov_amplitudes} and in Appendix \ref{appendix_u_and_v} is an excellent approximation in our experiment, this correction factor takes the simpler expression $X_{\bf q}(u_{\bf q}-v_{-\bf q})$ with $(u,v)$ the Bogoliubov amplitudes in the LP basis, and $X_{\bf q}$ is the usual (non-interacting) LP excitonic Hopfield coefficient. As a result, the phonon-condensate coupling rate which is proportional to $g_{xp}(\omega)^2$ ends up corrected by the factor $F_B=|u_{\bf q}-v_{-\bf q}|^2 \le 1$ by the Bogoliubov transformation of the elementary excitations.

Physically, $|u_{x,{\bf q}} - v_{x,-{\bf q}}|^2$ actually quantifies the fraction of excitonic density fluctuations in a Bogoliubov excitation, through which polaritons indeed couple to lattice deformations. Remarkably, for small momenta $|u_{x,{\bf q}} - v_{x,-{\bf q}}|^2 \ll 1$ and therefore the corresponding Bogoliubov excitations become effectively decoupled from the bath of thermal lattice phonons. This phenomenon is somewhat reminiscent of a superfluid behavior, in which, as a consequence of inter-particle interactions, the system effectively decouples from slowly-moving defects. In the context of reaching the \mr{regime dominated by photon vacuum fluctuations}, this decoupling favours its emergence in low momenta states, and thus increases the crossover temperature into the quantum regime as already pointed our in the previous section [cf. Eqs.~(\ref{Tcross_N},\ref{Tcross_G})].

Interestingly, we can derive a measurement of $F_{B}(\theta)$ from our experimental determination of $|v(\theta)/u(\theta)|^2$ presented in Section \ref{section_bogoliubov_amplitudes} as 
as $F_B^2=(1-|v/u|)/(1+|v/u|)$. The result is plotted in Fig.\ref{f4b}.c alongside the theoretical $F_B=|u_{x,{\bf q}}-v_{x,-{\bf q}}|^2 / X^2$. A Bogoliubov correction factor of $F_B\simeq 80\%$ is found at large angle and decreases to about $70\%$ at lower angles (for the data points which are the closest to the spectral filter), a trend which is in fair agreement with the theory, with deviations mostly coming from the weaker  measured  emission intensity of the ghost branch at low angles as compared to the theory.

The dataset presented in Fig.\ref{f4b}.d suggests the onset of a dipping behaviour of $F_B(\theta)$ at low angle with respect to higher angles. This dipping is an unambiguous signature of the Bogoliubov-mediated thermal protection mechanism from the phonon bath.

\section{Conclusion} 
\label{section_conclusion}

\mr{In this work we have shown that in general, a quantum fluid of polaritons embedded within a solid-state lattice has strongly modified characteristics and unique properties as compared to the same system considered in vacuum, or to a closed equilibrium system like ultracold atoms. Indeed, the open character of the system implies that the condensate interacts with the fluctuations of its surrounding environment. The environment constituted by the photon vacuum has been well investigated over the past decade, and shown to lead to striking properties that remain entirely relevant, such as the fact that this coupling produces a steady-state flux of Bogoliubov excitations, and that their subsequent emission is characterized by pairwise quantum correlations.

This work shows that the vibrating lattice that the condensate is embedded in constitutes a second environment that the condensate is coupled to. We show that it also produces a steady-state flux of Bogoliubov excitations, albeit with a very different spectral function, and a non-trivial lattice-temperature dependence. Strikingly, we predict a non-zero flux of elementary excitations even for a lattice temperature of $T=0$K.

We achieve a quantitative understanding of this contribution to the fundamental steady-state flux of elementary excitations via a quantitative comparison between experiment and theory. Our analysis shows in particular that in typical experimental conditions and temperatures, this fundamental steady-state flux is predominantly driven by thermal phonon fluctuations rather than by photon vacuum fluctuations. This is \mr{an obstacle if one wants to explore quantum correlations within the Bogoliubov excitation spectrum, which emerge when the Bogoliubov excitations are driven by photon vacuum fluctuations. Using our theoretical analysis, we propose realistic experimental strategies to overcome this difficulty in future experiments.} 

We also highlight a thermodynamical phenomenon by which the Bogoliubov transformation taking place within the quantum fluid reduces its coupling to the thermal bath of lattice phonons, and hence the amount of thermal energy that the condensate absorbs per unit of time in its steady-state. This thermal protection mechanism shows up in the steady-state in our system owing to its driven-dissipative nature. In equilibrium systems like ultracold atoms similar effects have been rather observed in the cooling or heating dynamics, in particular when the role of the heat bath is provided by impurities \cite{yan_2020, skou_2021} or by another species in the context of a two-components mixture \cite{Myatt_1997, Timmermans_1998, Tey_2010, Chen_2022}, such that the heat transfers are governed by an inelastic scattering mechanism similar to that of our exciton-phonon interaction Eq.(\ref{eq_Vxp}).}\\

\noindent\textit{Data availability.--}
The data to reproduce the figures presented in this paper are available at \url{https://zenodo.org/records/10972420}.

\begin{acknowledgments}
The authors wish to thank Iacopo Carusotto for insightful discussions. IF acknowledges support from the European Union Horizon's 2020 research and innovation programme under the Marie Sklodowska Curie grant agreement No 101031549 (QuoMoDys). This work was partly supported by the Paris Ile de France R\'egion in the framework of DIM SIRTEQ, by the RENATECH network and the General Council of Essonne, by the European Research Council via the project ARQADIA (949730) and the project ANAPOLIS (101054448), and by Centre of Quantum Technologies's Exploratory Initiative programme. AV acknowledges the European Union's Horizon 2020 research and innovation programme under the Marie Sklodowska-Curie grant agreement No 754303, and ERC grant LATIS.
\end{acknowledgments}

\appendix

\section{Details on the theoretical model}

\subsection{Input-output formalism}
\label{app_input_output}
In the experiment, we measure the emission intensity resolved in both frequency and wavevector, denoted as $I({\bf q},\omega)$, where ${\bf q}=(q_x,q_y)$ is the in-plane momentum. In order to compute it from the microscopic model, we adopt an input-output formalism. We work in the Heisenberg picture and make the standard assumption that the state is factorized at initial time $t=t_0$, namely $\hat\rho_0 = \hat \rho_p \otimes |0_{\rm phot}\rangle \langle 0_{\rm phot}|$, with $\hat\rho_p$ the intracavity polaritonic density matrix, and $|0_{\rm phot}\rangle$ the vacuum for the  extra-cavity photon modes. The latter is an excellent approximation in our experiment, as the typical frequencies involved ($\omega \sim 1.5 {\rm eV}$) in the problem correspond to a temperature $T>10^4{\rm K}\gg T_c$ with $T_c \sim 10{\rm K}$ the cryostat temperature, so that the extracavity modes can be safely considered in a vacuum state. We work in the Markovian approximation in which the bath retains no memory of its interaction with the system.

After a time interval $\Delta t=t-t_0$, the extacavity photon mode at $({\bf q}, k_z)$ contains a number of output photons given by $n({\bf q},k_z,\Delta t) = \langle \hat\alpha^\dagger_{{\bf q}, k_z}(t_0+\Delta t) \hat\alpha_{{\bf q}, k_z}(t_0+\Delta t) \rangle$, where the average is taken over the initial state, namely $\langle \dots \rangle = {\rm Tr}[\hat\rho_0 \dots]$. As per the experimental condition, the system is in its steady-state, so that this quantity does not depend on the initial time $t_0$. Experimentally, we measure the photon flux, per unit momentum and frequency, tunneling outside of the microcavity during a time interval $\Delta t$, which is macroscopic as compared to the system microscopic timescales, namely:
\begin{equation}
    I({\bf q},\omega) = \rho({\bf q},\omega) \lim_{\Delta t \rightarrow\infty} \left[ \frac{ n[{\bf q},k_z({\bf q},\omega), \Delta t] }{\Delta t}\right]
\end{equation}
where $\rho({\bf q},\omega) = \sum_{k_z} \delta(\omega - \omega_{{\bf q}, k_z}^{(\alpha)})$ is the partial density of states of extracavity photons, and $k_z({\bf q},\omega)=\sqrt{\omega^2/c^2-{\bf q}^2}$. 

Using Heisenberg equations of motion (EOM) for the extracavity photons $ i \hbar \partial_t\hat\alpha_{{\bf q}, k_z} = [\hat\alpha_{{\bf q}, k_z}(t), \hat{\cal V}_{\rm out}]$ where $\hat{\cal V}_{\rm out}$ is given in Eq.~\eqref{eq_V_photons} in the main text, we can relate their dynamical evolution to the one of intracavity photons as:
\begin{eqnarray}
    \hat\alpha_{{\bf q}, k_z}(t)  &=& e^{-i\omega_{{\bf q}, k_z}^{(\alpha)} (t-t_0)}\hat\alpha_{{\bf q}, k_z}(t_0)   \nonumber  \\
   && - i\kappa_{{\bf q}, k_z} \int_{t_0}^t dt' e^{-i\omega_{{\bf q}, k_z}^{(\alpha)}(t- t')} \hat a_{\bf q}(t')~.
    \label{eq_EOM_photons}
\end{eqnarray}
Since the photonic input state is the vacuum, only the second term in Eq.~\eqref{eq_EOM_photons} contributes to the output signal, leading to: 
\begin{eqnarray}
    I({\bf q},\omega) =   \lim_{\Delta t \to \infty} \frac{\gamma_{\rm cav}}{\pi\Delta t} \int_{t_0}^{t_0+\Delta t} dt_2 \int_{t_0}^{t_0+\Delta t} dt_1  \nonumber\\
     \times e^{-i\omega (t_2-t_1)} \langle \hat a_{\bf q}^\dagger(t_2) \hat a_{\bf q}(t_1) \rangle 
   \label{eq_output_signal_1_app}
\end{eqnarray}
with the cavity loss rate $ \gamma_{\rm cav} = \pi \rho({\bf q},\omega)\kappa_{{\bf q},k_z({\bf q},\omega)}^2$. Equation \eqref{eq_output_signal_1_app} thus shows that the photon emission intensity into the extracavity medium corresponds to the Fourier transform of the two-times correlations $\langle \hat a_{\bf q}^\dagger(t_2) \hat a_{\bf q}(t_1) \rangle$ of the intracavity photons, and hence of the polaritons. \mr{We further assume that $\Delta t$ is much larger than the inverse typical frequency range (namely, $\Delta t \gg 10^{-11}{\rm s}$ for frequencies in the meV range), allowing us to send $\Delta t  \rightarrow \infty$. In the steady-state, correlation functions depend only on the time difference $\tau=t_2-t_1$, and the output signal simplifies to:
\begin{equation}
    I({\bf q},\omega) =   \frac{\gamma_{\rm cav}}{\pi} \int_{-\infty}^{+\infty} d\tau  e^{-i\omega \tau} \langle \hat a_{\bf q}^\dagger(\tau) \hat a_{\bf q}(0) \rangle ~,
\end{equation}
namely it is proportional to the spectral function of the photonic component of the lower polaritons inside of the cavity.
}

\subsection{Solution of the mean-field steady-state equations}
\label{app-meanfield}
We provide here the details of the solution of Eqs.~(\ref{eq_Itp1}) and (\ref{eq_Itp2}).
Without loss of generality, we can take $\psi_x=\sqrt{n_x}$  and $\psi_c=\sqrt{n_c} e^{-i\phi}$.
Experimentally, we control the pump intensity $|F_p|^2$, and $n_x, n_c, \phi$ spontaneously take their steady-state values; but for the sake of deriving the solutions of the mean-field equations, we take $n_x$ as a parameter, and derive $n_c, \phi, |F_p|^2$ as a function of $n_x$. First, taking the real and imaginary parts of Eq.~\eqref{eq_Itp1}, we derive the relations:
\begin{eqnarray}
    (3g_s n_x / 2 - \Omega/2 )\sqrt{n_c} \cos\phi &=& \sqrt{n_x}(\omega_x + g_x n_x) \label{eq_cosphi} \\
    (g_s n_x / 2 - \Omega/2 )\sqrt{n_c} \sin\phi &=& -\sqrt{n_x} \gamma_x/2 \label{eq_sinphi}
\end{eqnarray}
Taking the square of these relations, and using $\cos^2\phi+\sin^2\phi=1$, we obtain the ratio:
\begin{equation}
    Q_n := \frac{n_c}{n_x} = \frac{(\omega_x + g_x n_x)^2}{(\Omega/2 - 3g_s n_x / 2)^2} + \frac{(\gamma_x/2)^2}{(\Omega/2 - g_s n_x / 2)^2}
\end{equation}
This relation gives $n_c$ as a function of $n_x$. Inserting $Q_n$ into Eqs.~\eqref{eq_cosphi} and \eqref{eq_sinphi}, we then find the relative phase $\phi$ as follows:
\begin{eqnarray}
    \cos\phi &=& \frac{\omega_x + g_x n_x}{\sqrt{Q_n}(3g_s n_x / 2 - \Omega/2)} \label{eq_cosphi_final} \\
    \sin\phi &=& \frac{\gamma_x/2}{\sqrt{Q_n}(\Omega / 2-g_s n_x / 2 )} \label{eq_sinphi_final}
\end{eqnarray}
In our experiment, $\Omega=3.28\,$meV is much larger than $g_s n_x=0.19\,$meV, such that $\cos \phi<0$ and $\sin \phi>0$. Using also $g_xn_x=0$ yields $\tan \phi\simeq \gamma_x/\omega_x(1/2-g_sn_x/\Omega)$ where $\gamma_x\sim 0.15\,$meV and $\omega_x=1.355\,$meV$/\hbar$ (we remind that the frequencies are expressed in the frame rotating at the laser frequency). We see that as long as $g_sn_x/\Omega \ll 1$, $\phi\simeq \pi$ where $\phi=\pi$ is the nominal phase difference in the linear regime for lower polariton states. Once $\phi$ and $n_c$ are obtained, we finally use Eq.~\eqref{eq_Itp2} to find $|F_p|^2$:
\begin{equation}
    |F_p|^2 = |(\omega_c-i\gamma_{\rm cav}/2)\sqrt{n_c} e^{-i\phi} + (\Omega/2-g_s n_x/2)\sqrt{n_x}|^2 ~.
\end{equation}
This provides the solution to the mean-field steady-state equations of the condensate.

\subsection{Polariton interactions in the Bogoliubov approximation}
\label{sec_app_interaction_bogo}
Here, we give the detailed expression for the polariton interaction terms in the Bogoliubov approximation. For the exciton-exciton interaction it results in:
\begin{equation}
    \hat{\cal V}_{xx} = (\hbar g_x n_x / 2) \sum_{\bf q} [\hat b_{{\bf q}_p + {\bf q}}^\dagger \hat b_{{\bf q}_p - {\bf q}}^\dagger + 4 \hat b_{{\bf q}_p + {\bf q}}^\dagger \hat b_{{\bf q}_p + {\bf q}} + \hat b_{{\bf q}_p + {\bf q}} \hat b_{{\bf q}_p - {\bf q}}] ~.
    \label{eq_Vxx_bogo}
\end{equation}
and for the saturation term:
\begin{eqnarray}
    \hat{\cal V}_{\rm sat} &= -(\hbar g_s / 2) \sum_{\bf q} \left( n_x[\hat a_{{\bf q}_p + {\bf q}}^\dagger \hat b_{{\bf q}_p - {\bf q}}^\dagger + 2 \hat a_{{\bf q}_p + {\bf q}}^\dagger \hat b_{{\bf q}_p + {\bf q}} + \right.\nonumber\\
    &\left.    \hat a_{{\bf q}_p + {\bf q}} \hat b_{{\bf q}_p - {\bf q}} + 2 \hat a_{{\bf q}_p + {\bf q}} \hat b^\dagger_{{\bf q}_p + {\bf q}}] + 
    \sqrt{n_c n_x}[e^{-i\phi} \hat b_{{\bf q}_p + {\bf q}}^\dagger \hat b_{{\bf q}_p - {\bf q}}^\dagger \right. \nonumber\\
    & \left. + (4 \cos\phi) \hat b_{{\bf q}_p + {\bf q}}^\dagger \hat b_{{\bf q}_p + {\bf q}} + e^{i\phi} \hat b_{{\bf q}_p + {\bf q}} \hat b_{{\bf q}_p - {\bf q}}]
    \right) ~.
    \label{eq_Vsat_bogo}
\end{eqnarray}
where $\phi$ is the relative phase between the excitonic and photonic condensate fields.

\subsection{Equations of motion for cavity photons}
\label{eom-phot}
The EOM for the cavity photons $\partial_t\hat a_{{\bf q}_p + {\bf q}} = (-i/\hbar)[\hat a_{{\bf q}_p + {\bf q}}(t), \hat{\cal H}_{\rm tot}]$ is made of three terms. A first term, contains the free part and the interaction between cavity photons and excitons [Eq.~\eqref{eq_H_pol}]:
\begin{equation}
    -\frac{i}{\hbar}[\hat a_{{\bf q}_p + {\bf q}}, \hat{\cal H}_{\rm pol}^{(0)}] = -i\omega_{c,{\bf q}_p + {\bf q}}^{(0)}\hat a_{{\bf q}_p + {\bf q}} - i\frac{\Omega}{2}\hat b_{{\bf q}_p + {\bf q}} ~,
    \label{eq_EOM_cavity_photons_1}
\end{equation}
a second term comes from the excitonic saturation mechanism (treated in the Bogoliubov approximation, as explained in the main text, Eq.~\eqref{eq_Vsat_bogo}):
\begin{equation}
    -\frac{i}{\hbar}[\hat a_{{\bf q}_p + {\bf q}}, \hat{\cal V}_{\rm sat}] = i(g_s n_x / 2) (\hat b^\dagger_{{\bf q}_p - {\bf q}} + 2 \hat b_{{\bf q}_p + {\bf q}})  ~,
    \label{eq_EOM_cavity_photons_2}
\end{equation}
and a third term containing the coupling to extra-cavity photons [Eq.~\eqref{eq_V_photons}]:
\begin{equation}
    -\frac{i}{\hbar}[\hat a_{\bf q}, \hat {\cal V}_{\rm out}] = -i \sum_{k_z} \kappa_{{\bf q}, k_z} \hat \alpha_{{\bf q}, k_z}~.
    \label{eq_EOM_cavity_photons_3}
\end{equation}

\subsection{Equations of motion for quantum well excitons}
\label{eom-exc}
The EOM for the quantum well excitons $\partial_t\hat b_{{\bf q}_p + {\bf q}} = (-i/\hbar)[\hat b_{{\bf q}_p + {\bf q}}(t), \hat{\cal H}_{\rm tot}]$ is made of three terms. A first term containing the free part and Rabi coupling to cavity photons:
\begin{equation}
    -\frac{i}{\hbar}[\hat b_{{\bf q}_p + {\bf q}}, \hat{\cal H}_{\rm pol}^{(0)}] = -i\omega_{x,{\bf q}_p + {\bf q}}^{(0)}\hat b_{{\bf q}_p + {\bf q}} - i\frac{\Omega}{2}\hat a_{{\bf q}_p + {\bf q}} ~,
    \label{eq_EOM_cavity_excitons_1}
\end{equation}
a second term comes from the excitonic saturation and Coulomb interactions (treated in the Bogoliubov approximation, Eqs.~\eqref{eq_Vsat_bogo} and \eqref{eq_Vxx_bogo}):
\begin{eqnarray}
    -\frac{i}{\hbar}[\hat b_{{\bf q}_p + {\bf q}}, \hat{\cal V}_{\rm sat} + \hat{\cal V}_{xx}] = i (g_s n_x / 2) (\hat a^\dagger_{{\bf q}_p - {\bf q}} + 2 \hat a_{{\bf q}_p + {\bf q}})  \nonumber\\
    + i g_s \sqrt{n_x n_x} [e^{-i\phi}\hat b^\dagger_{{\bf q}_p - {\bf q}} + (2\cos\phi) \hat b_{{\bf q}_p + {\bf q}}] 
        \nonumber\\
        - i g_x n_x (\hat b^\dagger_{{\bf q}_p - {\bf q}} + 2 \hat b_{{\bf q}_p + {\bf q}}) ~,~~~~~
\label{eq_EOM_cavity_excitons_2}
\end{eqnarray}
and a third term from the coupling to lattice phonons (in the Bogoliubov approximation, Eq.~\eqref{eq_Vxp_bogo}):
\begin{equation}
    -\frac{i}{\hbar}[\hat b_{{\bf q}_p + {\bf q}}, \hat {\cal V}_{xp}] =  \sqrt{n_x}\sum_{k_z} g_{xp}({\bf q}, k_z) (\hat c_{{\bf q}, k_z} - \hat c_{-{\bf q},k_z}^\dagger)~.
    \label{eq_EOM_cavity_excitons_3}
\end{equation}

\subsection{General results about the coupling to a bath}
\label{bath}
Cavity photons are coupled to a 3D continuum of extra-cavity photons, while excitons are coupled to a 3D continuum of lattice phonons. The coupling is respectively described by Eqs.~\eqref{eq_V_photons} and \eqref{eq_Vxp_bogo}. In both cases, it corresponds to a linear coupling to a bath of harmonic oscillators, namely a coupling which generically takes the form:
\begin{eqnarray}
    \hat{\cal H}_{\rm bath} / \hbar &=& \sum_{\bf q} [\hat a_{\bf q}^\dagger \sum_{k_z} (x_{{\bf q}, k_z} \hat \alpha_{{\bf q}, k_z} + y_{{\bf q}, k_z} \hat \alpha_{-{\bf q},k_z}^\dagger) + {\rm h.c.}] \nonumber\\
    && + \sum_{{\bf q}, k_z} \omega_{{\bf q}, k_z} \hat \alpha_{{\bf q}, k_z}^\dagger \hat \alpha_{{\bf q}, k_z}
    \label{eq_app_generic_bath_hamiltonian}
\end{eqnarray}
with $\hat a_{\bf q}$ either the cavity photon or cavity exciton, and $\hat \alpha_{{\bf q}, k_z}$ either the output photons or the lattice phonons. We now derive the contribution to the EOM of the $\hat a_{\bf q}$ operators resulting from such a generic coupling. We first write the EOM for the bath operators:
\begin{eqnarray}
    \partial_t \hat \alpha_{{\bf q}, k_z} &=& (-i/\hbar)[\hat \alpha_{{\bf q}, k_z}, \hat{\cal H}_{\rm bath}] \nonumber\\
    &=& -i[y_{-{\bf q},k_z} \hat a_{-\bf q}^\dagger + x_{{\bf q}, k_z}^* \hat a_{\bf q} + \omega_{{\bf q}, k_z} \hat \alpha_{{\bf q}, k_z}]
\end{eqnarray}
The formal solution reads:
\begin{eqnarray}
    \alpha_{{\bf q}, k_z}(t) e^{i\omega_{{\bf q}, k_z}t} &=& \alpha_{{\bf q}, k_z}(t_0) e^{i\omega_{{\bf q}, k_z}t_0} -i\int_{t_0}^t d\tau e^{i\omega_{{\bf q}, k_z}\tau}[\nonumber\\
    &&x_{{\bf q}, k_z}^* \hat a_{\bf q}(\tau) + y_{-{\bf q},k_z} \hat a_{-\bf q}^\dagger(\tau)] ~. \label{eq_solution_bath_EOM_app}
\end{eqnarray}
On the other hand, the contribution to the EOM for the system operators $\hat a_{\bf q}$ stemming from coupling to the bath reads:
\begin{equation}
    (-i/\hbar)[\hat a_{\bf q}(t), \hat{\cal H}_{\rm bath}] = -i\sum_{k_z}[x_{{\bf q}, k_z} \hat \alpha_{{\bf q}, k_z} + y_{{\bf q}, k_z} \hat \alpha^\dagger_{-{\bf q},k_z}] ~.
    \label{eq_app_system_EOM_bath}
\end{equation}
Injecting the formal solution of Eq.~\eqref{eq_solution_bath_EOM_app} for the bath operators into Eq.~\eqref{eq_solution_bath_EOM_app} yields a contribution to the EOM of the form:
\begin{eqnarray}
     &(-i/\hbar)[\hat a_{\bf q}(t), \hat{\cal H}_{\rm bath}] = \hat F_{\bf q}^{(x)}(t) - [\hat F_{-\bf q}^{(y)}(t)]^\dagger - \int_{-\infty}^\infty d\tau \{\nonumber\\
    &\Gamma_1(t-\tau) \hat a_{\bf q}(\tau) + \Gamma_2(t-\tau) \hat a_{-\bf q}^\dagger(\tau)\} ~. 
    \label{eq_EOM_generic_bath}
\end{eqnarray}
We have introduced the so-called Langevin forces:
\begin{equation}
    \hat F_{\bf q}^{(x)}(t) = -i\sum_{k_z} x_{{\bf q}, k_z} \hat{\alpha}_{{\bf q}, k_z}(t_0) e^{-i\omega_{{\bf q}, k_z}(t-t_0)} ~,
\end{equation}
and:
\begin{equation}
    \hat F_{\bf q}^{(y)}(t) = -i\sum_{k_z} y_{{\bf q}, k_z} \hat{\alpha}_{{\bf q}, k_z}(t_0) e^{-i\omega_{{\bf q}, k_z}(t-t_0)} ~.
\end{equation}
We have also introduced the damping kernels:
\begin{equation}
    \Gamma_1(t) = \Theta(t)\sum_{k_z}(|x_{{\bf q}, k_z}|^2 e^{-i\omega_{{\bf q}, k_z}t} + |y_{{\bf q}, k_z}|^2 e^{i\omega_{-{\bf q},k_z}t})
\end{equation}
and:
\begin{equation}
    \Gamma_2(t) =  \Theta(t)\sum_{k_z}(x_{{\bf q}, k_z} y_{-{\bf q},k_z} e^{-i\omega_{{\bf q}, k_z}t} + x_{-{\bf q},k_z} y_{{\bf q}, k_z} e^{i\omega_{-{\bf q},k_z}t})
\end{equation}
($\Theta$ is the Heaviside step function).
Notice that in the integral over the damping kernels, we have sent the initial time $t_0$ to $-\infty$, which amounts to assume that the support of the damping kernel (a few times the memory of the bath) is much smaller than $t-t_0$ (Markov approximation). 

The Langevin forces contributions ($\hat F$ terms in Eq.~\eqref{eq_EOM_generic_bath}) describe the forcing of the system by the external bath. The contribution on the second line of Eq.~\eqref{eq_EOM_generic_bath} ($\Gamma$ terms) describe the damping induced by the coupling to the bath, giving rise to a finite linewidth to Bogoliubov excitations.\\

\noindent\textit{Photonic bath.--} Specifying these results to our model, we obtain for cavity photons [cf. Eq.~\eqref{eq_EOM_cavity_photons_3}: $x_{{\bf q}, k_z} = \kappa_{{\bf q}, k_z}$]:
\begin{equation}
-\frac{i}{\hbar}[\hat a_{\bf q}(t), \hat{\cal V}_{\rm out}] = \hat F_{\bf q}(t) - \int_{-\infty}^\infty d\tau \gamma_{c,{\bf q}}(t-\tau) \hat a_{\bf q}(\tau)
\end{equation}
with the Langevin force:
\begin{equation}
    \hat F_{\bf q}(t) = -i\sum_{k_z} \kappa_{{\bf q}, k_z} \hat{\alpha}_{{\bf q}, k_z}^{(in)} e^{-i\omega_{{\bf q}, k_z}^{(\alpha)} t} 
    \label{eq_langevin_force_photons}
\end{equation}
with the input modes defined according to $\hat{\alpha}_{{\bf q}, k_z}^{(in)} := \hat{\alpha}_{{\bf q}, k_z}(t_0) e^{i\omega_{{\bf q}, k_z}^{(\alpha)}t_0}$, and with the damping kernel:
\begin{equation}
    \gamma_{c,{\bf q}}(t) = \Theta(t) \sum_{k_z}|\kappa_{{\bf q}, k_z}|^2 e^{-i\omega_{{\bf q}, k_z}^{(\alpha)}t} ~.
\end{equation}

\noindent\textit{Phononic bath.--} For cavity excitons which are coupled to the bath of thermal lattice solid-state phonons [cf. Eq.~\eqref{eq_EOM_cavity_excitons_3}: $x_{{\bf q}, k_z}=i\sqrt{n_x}g_{xp}({\bf q}, k_z)=-y_{{\bf q}, k_z}$], we find the contribution to EOM:
\begin{widetext}
\begin{equation}
    -\frac{i}{\hbar}[\hat b_{{\bf q}_p + {\bf q}}(t), \hat {\cal V}_{xp}] = \hat f_{\bf q}(t) - [\hat f_{-\bf q}(t)]^\dagger -\int_{-\infty}^\infty d\tau[\gamma_{x,{\bf q}}(t-\tau) - \gamma_{x,-{\bf q}}^*(t-\tau)][\hat b_{{\bf q}_p + {\bf q}}(\tau) + \hat b_{{\bf q}_p - {\bf q}}^\dagger(\tau)] ~,
    \label{eq_polaritons_EOM_phonons_1}
\end{equation}
\end{widetext}
where we have introduced the Langevin forces:
\begin{equation}
    \hat f_{\bf q}(t) = \sqrt{n_x}\sum_{k_z} g_{xp}({\bf q}, k_z) \hat c_{{\bf q}, k_z}^{(in)} e^{-i\omega_{{\bf q}, k_z}^{(\rm ph)}t}
    \label{eq_Langevin_phonons}
\end{equation}
with the input modes defined according to $\hat{c}_{{\bf q}, k_z}^{(in)} := \hat{c}_{{\bf q}, k_z}(t_0) e^{i\omega_{{\bf q}, k_z}^{(\rm ph)}t_0}$, as well as the damping kernels:
\begin{equation}
    \gamma_{x,{\bf q}}(t) = \Theta(t) \sum_{k_z} n_x g_{xp}^2({\bf q}, k_z) e^{-i\omega_{{\bf q}, k_z}^{(\rm ph)}t} ~.
\end{equation}

\subsection{Complete equations of motions in the frequency domain}
\label{eom-complete}
We shall gather all terms of the EOM and write them in the frequency domain.\\

\noindent\textit{Conventions for Fourier transforms.--} We first recall the notations for the Fourier transform of the operators:
\begin{eqnarray}
    &\hat O(\omega) = \int_{-\infty}^\infty dt e^{i\omega t} \hat O(t) \\
    &\hat O(t) = \frac{1}{2\pi} \int_{-\infty}^\infty d\omega e^{-i\omega t} \hat O(\omega) ~.
\end{eqnarray}
Notice that this convention implies the relation:
\begin{equation}
    \hat O^\dagger(\omega) = [\hat O(-\omega)]^\dagger ~.
\end{equation}

\noindent\textit{Photons Langevin forces.--}
The photon Langevin force Eq.~\eqref{eq_langevin_force_photons} reads in Fourier space:
\begin{eqnarray}
    \hat F_{\bf q}(\omega) &=& -i \sum_{k_z} \kappa_{{\bf q}, k_z} \hat\alpha_{{\bf q}, k_z}^{(in)} \int_{-\infty}^\infty dt e^{i(\omega - \omega_{{\bf q}, k_z}^{(\alpha)})t} \nonumber\\
    &=&  -i \sum_{k_z} \kappa_{{\bf q}, k_z} \hat\alpha_{{\bf q}, k_z}^{(in)} 2\pi\delta(\omega - \omega_{{\bf q}, k_z}^{(\alpha)})
\end{eqnarray}
Introducing the partial density of states $\rho_{{\bf q},\omega} = \sum_{k_z} \delta(\omega - \omega_{{\bf q}, k_z}^{(\alpha)})$, we rewrite the photon Langevin force as:
\begin{equation}
    \hat F_{\bf q}(\omega) =-i\kappa_{{\bf q},k_z(\omega)} 2\pi \rho_{{\bf q},\omega} \hat \alpha_{{\bf q},k_z(\omega)}^{(in)} ~,
    \label{eq_Langevin_photons_final}
\end{equation}
where $k_z(\omega)$ is such that $\omega_{{\bf q}, k_z}^{(\alpha)}=\omega$. It will be convenient to replace the label $k_z(\omega)$ by simply $\omega$.

\noindent\textit{Phonon Langevin forces.--}
The Langevin forces for the phonons, Eq.~\eqref{eq_Langevin_phonons} have a similar expression:
\begin{equation}
    \hat f_{\bf q}(\omega) = \sqrt{n_x} g_{xp}({\bf q},\omega) 2\pi \rho'_{{\bf q},\omega} \hat c_{{\bf q},\omega}^{(in)}
    \label{eq_Langevin_phonons_final}
\end{equation}
where the phonon partial density of states is $\rho'_{{\bf q},\omega} = \sum_{k_z} \delta(\omega - \omega_{{\bf q}, k_z}^{(\rm ph)})$, and we defined $\hat c_{{\bf q},\omega}^{(in)} := \hat c_{{\bf q},k'_z}^{(in)}$ where $k_z'$ is such that $\omega=\omega'_{{\bf q},k_z'}$. We defined similarly $g_{xp}({\bf q},\omega):= g_ {xp}({\bf q},k_z')$. Importantly, the density of states vanishes for $\omega<0$, and therefore $\hat f_{\bf q}^\dagger(\omega) = [\hat f_{\bf q}(-\omega)]^\dagger$ vanishes for $\omega>0$.\\

\noindent\textit{Polariton EOM in the frequency domain.--}
We are now ready to write down the complete EOM for the polaritons in the frequency domain. Gathering all contributions to the cavity photons and excitons EOM, we find:
\begin{widetext}
\begin{eqnarray}
    -i\omega \hat a_{{\bf q}_p + {\bf q}}(\omega) &=& -i\omega_{c,{\bf q}_p + {\bf q}}^{(0)} \hat a_{{\bf q}_p + {\bf q}}(\omega)  -i \frac{\Omega}{2} \hat b_{{\bf q}_p + {\bf q}}(\omega) \nonumber\\ 
    &&+ i(g_s n_x / 2) [\hat b^\dagger_{{\bf q}_p - {\bf q}}(\omega) + 2 \hat b_{{\bf q}_p + {\bf q}}(\omega)] \nonumber \\
    &&+ \hat F_{{\bf q}_p + {\bf q}}(\omega) -\gamma_{c,{\bf q}_p + {\bf q}}(\omega) \hat a_{{\bf q}_p + {\bf q}}(\omega)
\end{eqnarray}
and
\begin{eqnarray}
    -i\omega \hat b_{{\bf q}_p + {\bf q}}(\omega) &=& -i\omega_{x,{\bf q}_p + {\bf q}}^{(0)} \hat b_{{\bf q}_p + {\bf q}}(\omega)  -i \frac{\Omega}{2} \hat a_{{\bf q}_p + {\bf q}}(\omega) \nonumber\\ 
    &&+ i(g_s n_x / 2) [\hat a^\dagger_{{\bf q}_p - {\bf q}}(\omega) + 2 \hat a_{{\bf q}_p + {\bf q}}(\omega)]  \nonumber\\
    &&- i g_x n_x [\hat b^\dagger_{{\bf q}_p - {\bf q}}(\omega) + 2 \hat b_{{\bf q}_p + {\bf q}}(\omega)] \nonumber \\
    &&+ i g_s \sqrt{n_x n_c} [e^{-i\phi}\hat b^\dagger_{{\bf q}_p - {\bf q}}(\omega) + (2\cos\phi) \hat b_{{\bf q}_p + {\bf q}}(\omega)] \nonumber\\
    &&+ \hat f_{\bf q}(\omega) - \hat f^\dagger_{-\bf q}(\omega) - [\gamma_{x,{\bf q}}(\omega) - \gamma_{x,-{\bf q}}^*(\omega)][\hat b_{{\bf q}_p + {\bf q}}(\omega) + \hat b^\dagger_{{\bf q}_p - {\bf q}}(\omega) ]
\end{eqnarray}
Equivalently, we take the Hermitian conjugate of these last two equalities, use that $[\hat O(\omega)]^\dagger = \hat O^\dagger(-\omega)$, and change $(\omega,q)$ into $(-\omega, -q)$. We obtain:
\begin{eqnarray}
    -i\omega \hat a^\dagger_{{\bf q}_p - {\bf q}}(\omega) &=& i\omega_{c,{\bf q}_p - {\bf q}}^{(0)} \hat a^\dagger_{{\bf q}_p - {\bf q}}(\omega)  + i \frac{\Omega}{2} \hat b^\dagger_{{\bf q}_p - {\bf q}}(\omega) \nonumber\\ 
    &&- i(g_s n_x / 2) [\hat b_{{\bf q}_p + {\bf q}}(\omega) + 2 \hat b^\dagger_{{\bf q}_p - {\bf q}}(\omega)] \nonumber \\
    &&+ \hat F^\dagger_{{\bf q}_p - {\bf q}}(\omega) -\gamma^*_{c,{\bf q}_p - {\bf q}}(\omega) \hat a^\dagger_{{\bf q}_p - {\bf q}}(\omega)
\end{eqnarray}
and
\begin{eqnarray}
    -i\omega \hat b^\dagger_{{\bf q}_p - {\bf q}}(\omega) &=& i\omega_{X,{\bf q}_p - {\bf q}}^{(0)} \hat b^\dagger_{{\bf q}_p - {\bf q}}(\omega)  + i \frac{\Omega}{2} \hat a^\dagger_{{\bf q}_p - {\bf q}}(\omega) \nonumber\\ 
    &&- i(g_s n_x / 2) [\hat a_{{\bf q}_p + {\bf q}}(\omega) + 2 \hat a^\dagger_{{\bf q}_p - {\bf q}}(\omega)]  \nonumber\\
    &&+ i g_x n_x [\hat b_{{\bf q}_p + {\bf q}}(\omega) + 2 \hat b_{{\bf q}_p - {\bf q}}^\dagger(\omega)] \nonumber \\
    &&- i g_s \sqrt{n_x n_c} [e^{i\phi} \hat b_{{\bf q}_p + {\bf q}}(\omega) + (2\cos\phi) \hat b_{{\bf q}_p - {\bf q}}^\dagger(\omega)] \nonumber \\
    &&+ \hat f^\dagger_{-\bf q}(\omega) - \hat f_{\bf q}(\omega) - [\gamma^*_{x,-{\bf q}}(\omega) - \gamma_{x,{\bf q}}(\omega)][\hat b_{{\bf q}_p + {\bf q}}(\omega) + \hat b^\dagger_{{\bf q}_p - {\bf q}}(\omega) ]
\end{eqnarray}
We introduce the notations:
\begin{eqnarray}
    \tilde \gamma_{x,{\bf q}}(\omega) &=& \gamma_{x,{\bf q}}(\omega) - \gamma_{x,-{\bf q}}^*(\omega) \\
    \gamma_{\rm cav} &=& \gamma_{c,{\bf q}}(\omega) \\
    \mu_s &=&g_s n_x / 2 \\
    \mu_{sx} &=&g_x n_x - g_s \sqrt{n_x n_c} e^{-i\phi}
\end{eqnarray}
where the photonic loss rate $\gamma_{\rm cav}$ is taken as a real number (namely, we ignore any Lamb shift on the resonance frequencies that would stem from its imaginary part, and that is not observed in the experiment), and approximately frequency- and momentum-independent. The equations of motion can be finally summarized as:
\begin{equation}
    \begin{pmatrix}
        \hat a_{{\bf q}_p + {\bf q}}(\omega)\\
        \hat b_{{\bf q}_p + {\bf q}}(\omega)\\
        \hat a^\dagger_{{\bf q}_p - {\bf q}}(\omega)\\
        \hat b^\dagger_{{\bf q}_p - {\bf q}}(\omega)
    \end{pmatrix} = i[\omega\mathbf{1} - \tilde M_{\bf q}]^{-1}\begin{pmatrix}
        \hat F_{{\bf q}_p + {\bf q}}(\omega)\\
        \hat f_{\bf q}(\omega)  - \hat f^\dagger_{-\bf q}(\omega) \\
        \hat F^\dagger_{{\bf q}_p - {\bf q}}(\omega) \\ f^\dagger_{-\bf q}(\omega)- \hat f_{\bf q}(\omega)
    \end{pmatrix}
    \label{eq_input_output_final_app}
\end{equation}
where we introduced the $\tilde M_{\bf q}$ matrix:
\begin{equation}
   \tilde M_{\bf q} = \begin{pmatrix}
        \omega_{c,{\bf q}_p + {\bf q}}^{(0)}  - i\gamma_{\rm cav} & \Omega/2 - 2\mu_s & 0 & -\mu_s \\ 
        -2\mu_s +\Omega/2 & \omega_{x,{\bf q}_p + {\bf q}}^{(0)} + 2{\rm Re}(\mu_{sx})  - i\tilde \gamma_x & -\mu_s & \mu_{sx} - i\tilde\gamma_x \\
       0 & \mu_s & -\omega_{c,{\bf q}_p - {\bf q}}^{(0)} - i\gamma_{\rm cav} & -\Omega/2 + 2\mu_s \\
        \mu_s & -\mu_{sx}^* + i\tilde\gamma_x & 2\mu_s - \Omega/2 & -\omega_{x,{\bf q}_p - {\bf q}}^{(0)} - 2{\rm Re}(\mu_{sx}) + i\tilde \gamma_x 
    \end{pmatrix} ~.
    \label{eq_Mmat_app}
\end{equation}
\end{widetext}

\noindent\textit{Phenomenological damping terms.--} We notice that the exciton damping term $\tilde\gamma_x$ as predicted by the phononic bath model does not quantitatively accounts for the linewidth as measured experimentally. As a matter of fact, the finite linewidth of the excitons, $\gamma_{x}({\bf q})$ is caused by further decoherence mechanisms independent of the coupling to phonons, and we do not attempt to model it directly. Therefore, to faithfully model the experiment we neglect the $\tilde\gamma_x$ terms in the $\tilde M_{\bf q}$ matrix of Eq.~\eqref{eq_Mmat_app}, and add a phenomenological term $\gamma_{x}({\bf q})$, such that it results in a complete polaritonic linewidth $\gamma_{\bf q}$ that agree with the experiment. This leads to the expression of the $M_{\bf q}$ matrix as given by Eq.~\eqref{eq_M_mat} of the main text.\\

\mr{
\noindent\textit{Bogoliubov coefficients.--}
As mentioned in the main text, the $M_{\bf q}$ 
matrix is brought onto a diagonal form by $P_{\bf q} M_{\bf q} P_{\bf q}^{-1}$, with $P_{\bf q}$ containing the Bogoliubov coefficients $u, v$:
\begin{equation}
    P_{\bf q} = \begin{pmatrix}
      u_{lp,c,{\bf q}} & u_{lp,x,{\bf q}} & v_{lp,c,-{\bf q}} & v_{lp,x,-{\bf q}} \\
      u_{up,c,{\bf q}} & u_{up,x,{\bf q}} & v_{up,c,-{\bf q}} & v_{up,x,-{\bf q}} \\
      v_{lp,c,{\bf q}}^* & v_{lp,x,{\bf q}}^* & u_{lp,c,-{\bf q}}^* & u_{lp,x,-{\bf q}}^* \\
      v_{up,c,{\bf q}}^* & v_{up,x,{\bf q}}^* & u_{up,c,-{\bf q}}^* & u_{up,x,-{\bf q}}^*
    \end{pmatrix},
    \label{eq_4x4_bogoliubov_transformation}
\end{equation}
where the $\mathbf{q_p}$ dependence has been dropped for the sake of lighter notations. 
}

\subsection{Approximate decoupling between the upper and lower polariton}
\label{appendix_u_and_v}
In this section, we derive an approximate model assuming that the lower polariton (LP) and upper polariton (UP) do not hybridize under Bogoliubov transformation. This allows us to obtain analytical expressions for the frequencies and Bogoliubov coefficients which are in excellent agreement with the initial model.  Using the further approximation that the Bogoliubov coefficients are real, which we shall justify, we show that all four Bogoliubov coefficients for the LP can be reconstructed from the experimentally-measured ratio of ghost to normal branch signals. This allows us to reconstruct from the experimental data the quantity $|u_{x,{\bf q}} - v_{x,-{\bf q}}|^2$, characterizing the decoupling from lattice phonons, and discussed in section \ref{FB_section}. Our starting point is the matrix $M_{\bf q}$ [Eq.~\eqref{eq_M_mat} of the main text] expressed as:
\begin{equation}
    M_{\bf q} = \begin{pmatrix} A_{\bf q} & B \\ -B^* & -A_{-\bf q} \end{pmatrix}
\end{equation}
 with:
\begin{equation}
    A_{\bf q} = A_{\bf q}^* =A_{\bf q}^t = \begin{pmatrix}
      \omega_{c,{\bf q}_p + {\bf q}}^{(0)} & \Omega/2 - 2\mu_s \\ 
        \Omega/2 -2\mu_s & \omega_{x,{\bf q}_p + {\bf q}}^{(0)} + 2{\rm Re}(\mu_{sx}) 
    \end{pmatrix}
\end{equation}
and:
\begin{equation}
    B = B^t = \begin{pmatrix}
      0 & -\mu_s \\ 
      -\mu_s & \mu_{sx}
    \end{pmatrix}
\end{equation}
For the sake of notation, we omit ${\bf q}_p$ in the labels; in all expressions, one should make the substitution $\pm {\bf q} \to {\bf q}_p \pm {\bf q}$. We first perform a unitary transformation to the (non-interacting) polariton basis. We introduce the $2\times2$ unitary matrix $U_0$ that diagonalizes the non-interacting problem ($\mu_s=\mu_{sx}=0$), namely:
\begin{equation}
    U_{0,{\bf q}} = \begin{pmatrix}
      C_{\bf q} & -X_{\bf q} \\
      X_{\bf q} & C_{\bf q}
    \end{pmatrix}
\end{equation}
with:
\begin{equation}
  U_{0,{\bf q}}^t \begin{pmatrix}
      \omega_{c,{\bf q}}^{(0)} & \Omega/2 \\ 
        \Omega/2 & \omega_{x,{\bf q}}^{(0)}
    \end{pmatrix} U_{0,{\bf q}} = \begin{pmatrix}
       \omega_{lp,{\bf q}}^{(0)} & 0 \\
       0 & \omega_{up,{\bf q}}^{(0)}
    \end{pmatrix}
\end{equation}
with the non-interacting resonance frequencies:
\begin{equation}
    \omega_{lp/up,{\bf q}}^{(0)} = \frac{\omega_{c,{\bf q}}^{(0)} + \omega_{x,{\bf q}}^{(0)}}{2} \mp \frac12 \sqrt{
        [\omega_{c,{\bf q}}^{(0)} - \omega_{x,{\bf q}}^{(0)}]^2 + \Omega^2
    } ~.
\end{equation}
The so-called Hopfield coefficients $X_{\bf q}$ and $C_{\bf q}$ are real, and define the lower and upper polariton operators via:
\begin{eqnarray}
    \hat a_{lp,{\bf q}} &=& C_{\bf q} \hat a_{\bf q} + X_{\bf q} \hat b_{\bf q} \\
    \hat a_{up,{\bf q}} &=& -X_{\bf q} \hat a_{\bf q} + C_{\bf q} \hat b_{\bf q} 
\end{eqnarray}
We then write the $M_{\bf q}$ matrix in this basis, namely:
\begin{eqnarray}
    M'_{\bf q} &=& \begin{pmatrix}
      U_{0,{\bf q}}^t & 0 \\
      0 & U_{0,-{\bf q}}^t
    \end{pmatrix} M_{\bf q} \begin{pmatrix}
      U_{0,{\bf q}} & 0 \\
      0 & U_{0,-{\bf q}}
    \end{pmatrix} \nonumber\\
    &=& \begin{pmatrix}
        A_{\bf q}' & B_{\bf q}' \\
        -(B'_{-\bf q})^* & -A'_{-\bf q}
    \end{pmatrix}
\end{eqnarray}
We have to evaluate the matrices $A_{\bf q}' = U_{0,{\bf q}}^t A_{\bf q} U_{0,{\bf q}}$ and $B'_{\bf q} = U_{0,{\bf q}}^t B U_{0,-{\bf q}}$. The final result is:
\begin{equation}
    A_{\bf q}' = \begin{pmatrix}
      \omega_{lp,{\bf q}}^{(0)} + 2{\rm Re} (gn)_{{\rm LL},{\bf q},{\bf q}} & 2{\rm Re} (gn)_{{\rm LU},{\bf q},{\bf q}} \\
      2{\rm Re} (gn)_{{\rm LU},{\bf q},{\bf q}} & \omega_{up,{\bf q}}^{(0)} + 2{\rm Re} (gn)_{{\rm UU},{\bf q},{\bf q}}
    \end{pmatrix}
\end{equation}
\begin{equation}
    B_{\bf q}' = \begin{pmatrix}
      (gn)_{{\rm LL},{\bf q},-{\bf q}} & (gn)_{{\rm LU},{\bf q},-{\bf q}} \\
      (gn)_{{\rm LU},{\bf q},-{\bf q}} & (gn)_{{\rm UU},{\bf q},-{\bf q}}
    \end{pmatrix} ~,
\end{equation}
where we define the ($q$-dependent) effective interactions as:
\begin{eqnarray*}
    (gn)_{{\rm LL},{\bf q},{\bf q}'} &=& -(C_{\bf q} X_{{\bf q}'} + X_{\bf q} C_{{\bf q}'}) \mu_s + X_{\bf q} X_{{\bf q}'} \mu_{sx} \\
    (gn)_{{\rm UU},{\bf q},{\bf q}'} &=& (C_{\bf q} X_{{\bf q}'} + X_{\bf q} C_{{\bf q}'})\mu_s + C_{\bf q} C_{{\bf q}'} \mu_{sx} \\
    (gn)_{{\rm LU},{\bf q},{\bf q}'} &=& (X_{\bf q} X_{{\bf q}'} - C_{\bf q} C_{{\bf q}'}) \mu_s + X_{\bf q} C_{{\bf q}'} \mu_{sx}
\end{eqnarray*}

We now make the simplification that the upper- and lower-polaritons do not hybridize, namely we set $(gn)_{\rm LU} = 0$. The $M'_{\bf q}$ matrix then splits into two independent $2\times 2$ matrices, describing independent  Bogoliubov transformations for the upper- and lower-polaritons. We have verified that, in the conditions of the experiment, the coefficients of the ($2 \times 2$) Bogoliubov transformations found in this approximation are almost identical to the exact ($4 \times 4$) ones. We then have two matrices do diagonalize:
\begin{equation}
    M'_{lp,{\bf q}} = \begin{pmatrix}
        a_{\bf q} & b_{\bf q} \\
        -b_{-\bf q}^* & -a_{-\bf q}
    \end{pmatrix}
    \label{eq_Mprime_LP}
\end{equation}
with:
\begin{eqnarray}
    a_{\bf q} &:=& \omega_{lp,{\bf q}}^{(0)} + 2{\rm Re} (gn)_{{\rm LL},{\bf q},{\bf q}} \\
    b_{-\bf q} &:=& (gn)_{{\rm LL},{\bf q},-{\bf q}}
\end{eqnarray}
for the LP, and similarly for the UP with $a_{\bf q} = \omega_{up,{\bf q}}^{(0)} + 2{\rm Re} (gn)_{{\rm UU},{\bf q},{\bf q}}$ and $b_{\bf q} = (gn)_{{\rm UU},{\bf q},-{\bf q}}$. The diagonalization is of the form:
\begin{equation}
    M'_{lp,{\bf q}} = \begin{pmatrix}
        u_{\bf q}^* & -v_{\bf q} \\
        -v_{-\bf q}^* & u_{-\bf q}
    \end{pmatrix} \begin{pmatrix}
       \omega_{lp,{\bf q}} & 0 \\
        0 & -\omega_{lp,-{\bf q}}
    \end{pmatrix} \begin{pmatrix}
        u_{\bf q} & v_{-\bf q} \\
        v_{\bf q}^* & u_{-\bf q}^*
    \end{pmatrix}
\label{eq_Mprime_LP_conventions}
\end{equation}
where we introduce the $u$ and $v$ Bogoliubov coefficients associated to the lower polariton. As we shall only be interested in the lower-polariton $u,v$ coefficients, we omit their $lp$ labels. $\omega_{lp,{\bf q}}$ gives the LP dispersion relation in the presence of interactions. Similarly, for the UP the eigenvalues of $M'_{up,{\bf q}}$, namely $\omega_{up,{\bf q}}$ and $-\omega_{up,-{\bf q}}$, give the UP dispersion relation in the presence of interactions. Given that Re$(gn)_{\rm UU} \ll \omega_{up,{\bf q}}^{(0)}$ for all relevant ${\bf q}$'s, the matrix $M'_{up,{\bf q}}$ is almost diagonal, and therefore the Bogoliubov eigenmodes are almost identical to the non-interacting upper polaritons (the Bogoliubov transformation diagonalizing $M'_{up,{\bf q}}$ is close to the identity, something which we verified numerically).

We now explicitly proceed to the diagonalization of the $M'_{lp,{\bf q}}$ matrix in Eq.~\eqref{eq_Mprime_LP}. Notice that $b_{-\bf q}=b_{\bf q}$. We then write:
\begin{equation}
    M'_{lp,{\bf q}} = \frac{a_{\bf q} - a_{-\bf q}}{2} \mathbf{1} + \begin{pmatrix}
        \bar a_{\bf q} & b_{\bf q} \\
        -b_{\bf q}^* & -\bar a_{\bf q}
    \end{pmatrix} 
\end{equation}
with $\bar a_{\bf q} = (a_{\bf q} + a_{-\bf q}) / 2$. The eigenvalues of $M'_{lp,{\bf q}}$ are then $\omega_{lp,{\bf q}}$ and $-\omega_{lp,-{\bf q}}$ with:
\begin{equation}
    \omega_{lp,{\bf q}} = \frac{a_{\bf q} - a_{-\bf q}}{2} + \sqrt{
        \bar a_{\bf q}^2 - |b_{\bf q}|^2
    }
\end{equation}
We introduce the notation $\tilde \omega_{\bf q} = \sqrt{\bar a_{\bf q}^2 - |b_{\bf q}|^2}$. Notice that we focus on the regime where $\bar a_{\bf q}^2 > |b_{\bf q}|^2$, which is always the case in the regime accessible to our measurements. Otherwise, the eigenvalues have an imaginary part, and we find that $|u_{\bf q}|^2=|v_{\bf q}|^2$, \mr{so that the eigenmodes do not describe proper bosonic excitations. This latter case corresponds to the flat part in the excitation spectrum,} which shall not be discussed further in this paper. Solving the eigenvalue equations using the conventions of Eq.~\eqref{eq_Mprime_LP_conventions}, the convention that $u_{\bf q}$ is real, and the normalization condition $|u_{\bf q}|^2-|v_{\bf q}|^2=1$, we find the coefficients:
\begin{eqnarray}
    u_{\bf q} = u_{-\bf q} &=& \sqrt{\frac{\bar a_{\bf q}}{2\tilde \omega_{\bf q}} + \frac{1}{2}} \\
    v_{\bf q} = v_{-\bf q} &=& \frac{b_{\bf q}}{|b_{\bf q}|}\sqrt{\frac{\bar a_{\bf q}}{2\tilde \omega_{\bf q}} - \frac{1}{2}}
\end{eqnarray}
Notice in particular the the $v_{\bf q}$ coefficient is complex, with a phase given by the phase of $b_{\bf q}=(gn)_{{\rm LL},{\bf q},-{\bf q}}$. In practice, though, this phase is negligible, and we take $v_{\bf q}$ real.

The Bogoliubov eigenmode of the lower polariton is described by the annihilation operator (recall that a non-zero pump momentum ${\bf q}_p$ should be re-introduced in these expressions: $\pm {\bf q} \to {\bf q}_p \pm {\bf q}$):
\begin{eqnarray}
    &\hat \beta_{lp,{\bf q}} = u_{\bf q} \hat a_{lp,{\bf q}} + v_{-\bf q} \hat a_{lp,-{\bf q}}^{\dagger} \\
    &= u_{\bf q}(C_{\bf q} \hat a_{\bf q} + X_{\bf q} \hat b_{\bf q}) + v_{-\bf q}(C_{-\bf q} \hat a_{-\bf q}^\dagger + X_{-\bf q} \hat b_{-\bf q}^\dagger)
\end{eqnarray}
We therefore have the identification:
\begin{eqnarray}
    u_{lp,c,{\bf q}} &=& u_{\bf q} C_{\bf q} \nonumber\\ 
    u_{lp,x,{\bf q}} &=& u_{\bf q} X_{\bf q} \nonumber\\
    v_{lp,c,-{\bf q}} &=& v_{-\bf q} C_{-\bf q} \nonumber\\  
    v_{lp,x,-{\bf q}} &=& v_{-\bf q} X_{-\bf q}
    \label{eq_uv_lp_approx}
\end{eqnarray}
Therefore, the experimental measurement of $v_{lp,c,-q} / u_{lp,c,q}$ gives direct access to the ratio $v_{-\bf q}/u_{\bf q}$ of the lower-polariton Bogoliubov transformation. Assuming that $v_{\bf q}=v_{-\bf q}$ is real (an excellent approximation in the regime of the experiment), and using the normalization $u_{\bf q}^2 - v_{\bf q}^2 = 1$ \mr{valid in the non-flat part of the dispersion relation}, we then reconstruct both $u_{\bf q}$ and $v_{\bf q}$. Using the knowledge of the Hopfield coefficients $X_{\bf q}, C_{\bf q}$ of the non-interacting problem, we may then reconstruct all four Bogoliubov coefficients $u_{c,{\bf q}}, u_{x,{\bf q}}, v_{c,{\bf q}}, v_{x,{\bf q}}$, and in particular the thermal decoupling coefficient $|u_{x,{\bf q}} - v_{x,-{\bf q}}|^2$, which is discussed in section \ref{FB_section}. The quantitative agreement further confirms both the validity of the model, as well as the approximate decoupling between LP and UP as discussed in this section.


\subsection{Estimating the exciton-phonon interaction $n_x\gamma_{xp}(\omega)$ }
\label{app_gamma_xp}
In this Appendix, we provide further details and quantitative estimates for the exciton-phonon interaction strength $\gamma_{xp}(\omega)$ introduced in Eq.~\eqref{eq_gamma_xp} of the main text.
In the limit of strong confinement of excitons in the $z$-direction of the quantum well, the exciton-phonon coupling amplitude is given by:
\begin{equation}
\hbar g_{xp}({\bf q}, k_z) = \sqrt{\frac{\hbar\sqrt{{\bf q}^2 + k_z^2}}{2 \rho V v_s}} (V_e I_e^p I_e^z - V_h I_h^p I_h^z)
\label{gxp1}
\end{equation}
where $\rho=5.3\times 10^3\,$kg.m$^{-3}$ is GaAs density, $v_s=4.7\times 10^3\,$m.s$^{-1}$ is the longitudinal acoustic speed of sound in GaAs, $V$ is a quantization volume, $V_{e,h}=(-7,2.7)\,$eV are the deformation potentials in GaAs for band-edge electrons and holes respectively.
\begin{equation}
    I_e^z \simeq I_h^z \simeq \exp(-k_z^2/q_{z,{\rm cut}}^2) \\
\end{equation}
is a Gaussian approximation of the $k_z$-component of the electron and hole envelope wavefunction which is assumed to be mostly determined by the quantum well thickness $L_z$, imposing the cutoff wavevector $q_{z,{\rm cut}}\approx 0.9\times 2\pi/L_z$.
\begin{equation}
I_{e(h)}^p = [1 + (m_{e(h)}/(2M) q a_B)^2]^{-3/2}
\end{equation}
is the in-plane component of the electron-hole wavefunction resulting from its bound state character of Bohr radius $a_B\simeq 10\,$nm. $m_{e,h}$ are the electron and hole effective masses and $M=m_e+m_h$ is the excitonic mass. Notice that $g_{xp}({\bf q}, k_z)=g_{xp}({\bf q}, -k_z)$, which allows us to replace unambiguously $k_z$ by $\omega$ using the dispersion relation $\omega=v_s\sqrt{{\bf q}^2+k_z^2}$. In the parameter regime of the experiment, $q \ll k_z$ so that $\omega \approx v_s |k_z|$. Furthermore, $[m_{e(h)}/(2M) q a_B]^2 \ll 1$. Therefore the coupling simplifies to:
\begin{equation}
    \hbar g_{xp}(\omega) = \sqrt{\frac{\hbar \omega}{2 \rho V v_s^2}} (V_e - V_h) \exp(-\frac{\omega^2}{v_s^2q_{z,{\rm cut}}^2}) ~.
\end{equation}
An explicit expression for $\gamma_{xp}(\omega)$ in Eq.~\eqref{eq_gamma_xp} can now be determined. We define the excitonic density in the quantum well as $n_x=N_x/A$ where $A$ is the condensate area and $V=AL_z$. The density of states $ \rho'_{{\bf q},\omega}$ counts the number of phonons wavevectors $k_z$ matching the condition $\omega = \omega_{{\bf q}, k_z}^{(\rm ph)} = v_s\sqrt{{\bf q}^2 + k_z^2}$, namely it is such that:
\begin{equation}
    \rho'_{{\bf q},\omega} d\omega = (L_z/2\pi) dk_z.
\end{equation}
Using the fact that in our experiment $v_s |{\bf q}| \ll \omega$, we get:
\begin{equation}
\rho'_{{\bf q},\omega} \approx \Theta(\omega) \frac{L_z}{2\pi v_s} 
\end{equation}
which is approximately constant ($\Theta$ is the Heaviside step function). We have further verified numerically the validity of this approximation. So finally,
\begin{equation}
\gamma_{xp}(\omega)n_x = \frac{n_xL_zA}{2v_s}\Theta(\omega)g_{xp}(\omega)^2 ~,
\end{equation}
that can be more explicitly given as 
\begin{equation}
\gamma_{xp}(\omega)n_x= \frac{\omega n_x}{4v_s^3 \rho \hbar} (V_e - V_h)^2\exp{\left(-\frac{2\omega^2}{v_s^2q_{z,{\rm cut}}^2}\right)}.
\end{equation}
Using this expression to fit the decay of integrated intensity versus angle for different temperatures in the vanishing interaction regime, and relying on the fact in this regime of very low laser excitation, we have a temperature measurement in the cryostat that provides a reliable measurement the phonon temperature in the microcavity (the temperature probe is glued next to the microcavity in the same way), we could determine that $q_{z,{\rm cut}}$ is best fitted with $L_z=8.5\,$nm instead of the nominal quantum well thickness $L_z=17\,$nm. A good reason for this mismatch is that the strong confinement assumption $a_B>L_z$, with $a_b=10\,$nm, is in fact not well checked in our thick quantum well. In the weaker confinement regime where $a_B<L_z$, the contribution of the bound electron-hole wavefunction contributes as the shortest length scale and hence contributes more to $q_{z,{\rm cut}}^2$ than $L_z$. The apparent $L_z$ in the strong confinement description is thus expected to decrease by a factor of the order of $\sim L_z/a_B$. Another contribution to this reduced $L_z$ is that the the expression $q_{z,{\rm cut}}\sim 0.9\times 2\pi/L_z$ is obtained by fitting the exact wavefunction in a finite height quantum well with a vertical transition edge between the barriers and the quantum well, while in reality, the transition edges are typically much smoother than that due to Indium diffusion, that can result in tighter confinement length along $z$ for the ground state (see e.g. \cite{Muraki_1993}). 


\section{Main experimental parameters and physical quantities}
\mr{In this section, we provide a quantitative summary of the physical parameters entering the modelling of our experiment, and of the key physical quantities introduced in the article.}
\begin{table*}
\caption{\label{param_tab} Main system parameters and physical quantities.}
\mr{\begin{tabular}{|l|l|l|} 
\hline
\textbf{Notation} & \textbf{Description} & \textbf{Value}\\ 
\hline
\hline
$\hbar\omega_{{\rm cav}}(\mathbf{q}=0)$& Cavity mode energy & $1450.54\,$meV \\ 
\hline
$\hbar\omega_{x}$& Excitonic transition energy & $1450.36\,$meV \\ 
\hline
$\hbar\omega_{lp}(\mathbf{q}=0)$ & Lower polariton ground state energy & $1448.81\,$meV\\
\hline
$\hbar\omega_{\rm las}$& laser to polariton mode detuning & $\hbar\omega_{lp}(\mathbf{q}=0)+0.19\,$meV\\ 
\hline 
$\hbar\Omega$ & Rabi splitting & $3.28\,$meV\\
\hline
$\hbar\gamma_{x,0}$ & Excitonic transition linewidth & $\simeq 0.12\,$meV \\ 
\hline
$\beta$& Momentum dependence parameter of excitonic linewidth & $\simeq 0.045\,$meV.$\mu$m$^2$ \\ 
\hline
$\hbar\gamma_{{\rm cav}}$& Cavity mode linewidth & $\simeq 0.025\,$meV \\ 
\hline
$\hbar \gamma_{xp}^{M}$& Max value (over $\mathbf{q}$) of the exciton-phonon interaction strength& $\simeq 5\times 10^{-3}$meV.$\mu$m$^2$ \\ 
\hline
$n_x$& excitonic density  & $[0.32,3.2]\times 10^{11}$ cm$^{-2}$\\ 
\hline
$\hbar g_sn_x$& Interaction energy: saturation contribution  & $0.19\,$meV\\ 
\hline
$\hbar g_xn_x$& Interaction energy: Coulomb contribution  & $0\,(-0;+0.3)\,$meV\\ 
\hline
$T_c$& Temperature: cryostat & $[6,11]\,$K \\ 
\hline
$T$& Temperature: thermal phonons bath & $[6,15]\,$K \\ 
\hline
$T_{GN}$& For $T\gg T_{GN}$, the Normal-Ghost brightness ratio is T-independent & $\simeq 15\,$K \\ 
\hline
$T_{EB}$& Charac. Temp.: equal Normal and Ghost mode brightness  & $\simeq 5.2\,$K\\ 
\hline
$T^{(cr)}_{N,G}(\mathbf{q})$& Crossover Temp.: phonon bath to photon vacuum dominated emission & $\gtrsim 1\,$K\\ 
\hline
\end{tabular}}
\end{table*}


\pagebreak

\onecolumngrid
\begin{center}
  \textbf{\large SUPPLEMENTARY INFORMATION \\ Spontaneous emission of Bogoliubov excitations by a quantum fluid of light coupled to thermal solid-state phonons}\\[1cm]
\end{center}
\twocolumngrid

\section {Hysteresis characterization at $T_c=6.6\,$K}
\label{app_hysteresis}

\begin{figure}[bt]
\includegraphics[width=0.6\columnwidth]{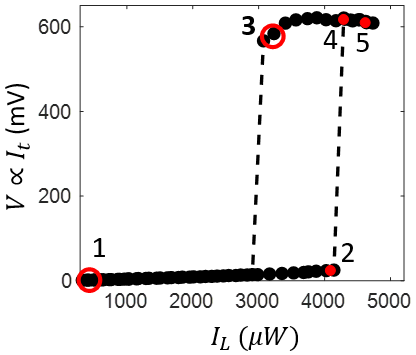}
\caption{Measured (black round symbol) transmitted intensity $I_t(\omega_p,k_p)^2$ at the frequency $\omega_p$ and momentum $k_p$ of the condensate, as a function of the laser intensity $I_L$. The dashed lines highlight the edges of the bistable region. the open symbols labeled '1' and '3' correspond to the two working points where the spectral function shown in Fig.2.a and Fig.2.c of the main text has been measured. The spectral functions at the working points shown by the red filled symbols have also been measured and are shown in Fig.\ref{fA3}.}  
\label{fA1}
\end{figure}

The measurement of the transmitted laser intensity $I_t$ ($I_t\propto V$, where $V$ is the voltage at the output of an amplified photodiode) versus the laser impinging intensity $I_L$ (measured before the sample with a calibrated powermeter) at $T_c=6.6\,$K is shown in Fig.\ref{fA1}. It exhibits a hysteretic behaviour as expected in the regime where the addressed polariton state frequency (expressed in the frame rotating at the laser frequency $\omega_p$) is negative and $-\omega_{lp}(k_p)/\gamma_{lp}\simeq 5>\sqrt{3}/2$. A large bistable region is found for $I_L\in [3.2,4]\,$mW. 

The state labelled '1' in Fig.\ref{fA1}, corresponding to $I_L=0.4\,$mW, is the non-interacting regime measured in Fig.2.a of the main text and analyzed in the main text. The corresponding intracavity polariton condensate density is proportional the transmitted intensity $I_{t,1}\propto 1.36\,$mV. Note that at this position in the hysteresis curve $I_t(I_L)$, the detuning between the laser and the excited polariton state (cf. $-\omega_{lp}(k_p)/\gamma_{lp}$) is significant such that the laser is poorly coupled to the excited polariton state.

The so-called interacting regime analyzed in the main text is the working point labelled '3' in Fig.\ref{fA1}. It corresponds to $I_L=3.21\,$mW, and is situated in the upper state of the bistable system, in which the nonlinearity driven blueshift essentially cancels the detuning between the polariton state and the laser drive. The coupling between the two is thus optimal and the condensate density more than 2 orders of magnitude higher, with $I_{t,3}\propto 577\,$mV. The two-body interaction energy ratio between state '3' and '1' can be derived quantitatively from this measurement as $(g_sn_{x,3})/(g_sn_{x,1})=I_{t,3}/I_{t,1}=425$, where $g_sn_{x,3}=0.19\,$meV is found in the main text analysis.

Working point '3' has been chosen for the interacting regime at the expense of points at higher $I_L$ such as point '4'and '5', as their corresponding interaction energies $g_sn_x\propto I_t$ exhibit only a very small increase with respect to working point 3 that comes with some degree of spectral broadening that typically increases the experimental uncertainty of the observable we are interested in in this work. 

\section {Data analysis methods}
\label{app_data_analysis}

\begin{figure*}[bt]
\includegraphics[width=0.95\textwidth]{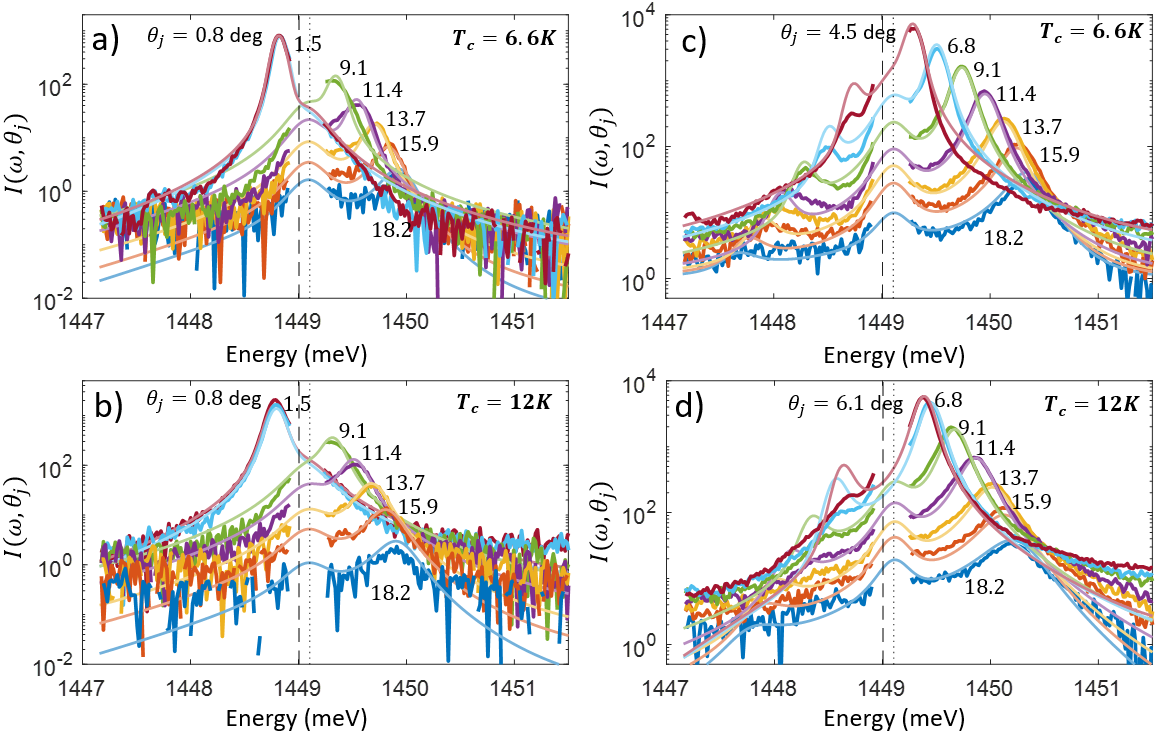}
\caption{Spectra extracted from the spectral function at the selected angles mentioned in each panels for vanishing interaction energy at (a) $T_c=6.6\,$K, and (b) $T_c=12\,$K (b), and for (c) $g_sn_x=0.19\,$meV and $T=6.6\,$K, and for (d) $g_sn_x=0.22\,$meV and $T_c=12\,$K. The measured and calculated spectra are plotted as thick  and thin lines respectively. The dashed vertical line highlights the laser energy, and the dotted one highlights the energy of the quasi-resonant extra peak discussed in section \ref{section_data_analysis}}.
\label{fA2}
\end{figure*}

\begin{figure*}[hbt]
\includegraphics[width=\textwidth]{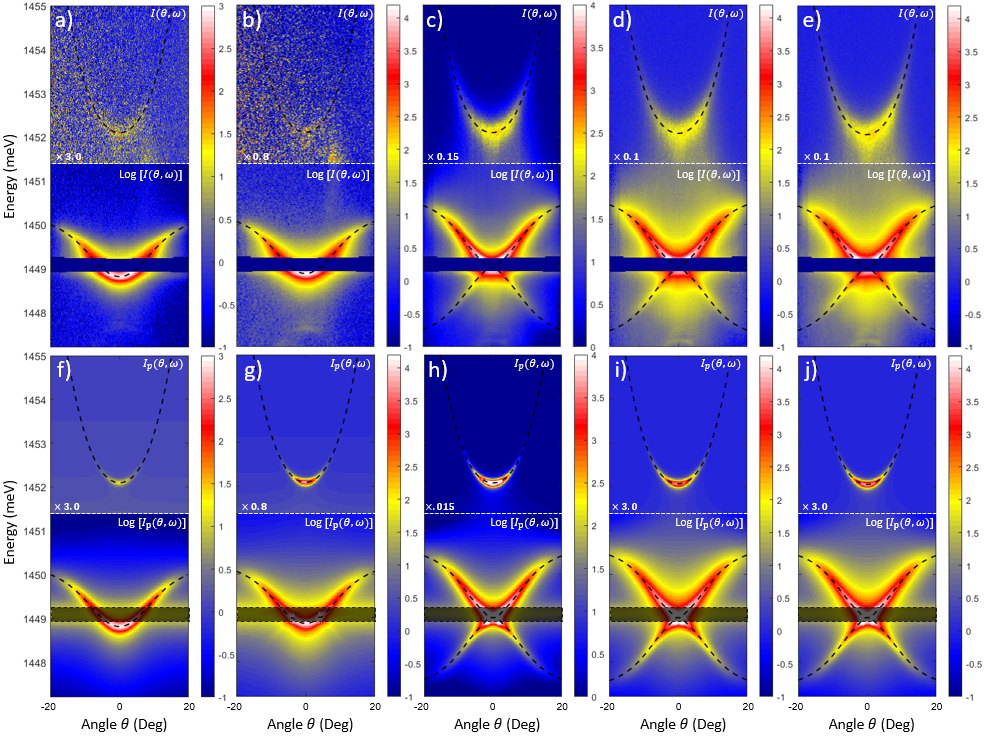}
\caption{Measured (a-e) and calculated (f-j) spectral functions (color scaled) for the 5 different States labeled '1' to '5' in Fig.\ref{fA1}. Both for the measurement and the theory, $I_{p}$ (involving only the phonon bath), we use a Log scale between $1447.1\,$meV and $1451.3\,$meV (lower polariton area) and a linear scale above (upper polariton area). The black dashed lines in all panels is calculated dispersion relations for three Bogoliubov modes (lower normal and ghost polariton branch and upper polariton branch). The fit parameters can be found in the SI text. The white multiplying factors have been applied to the emission intensity of the upper polariton region for better color contrast.}
\label{fA3}
\end{figure*}

\subsection{Extracting relevant observables from raw measurement of $I(\theta,\omega)$}
\label{section_data_analysis}
In order to extract a measurement of the dispersion relation $\omega_j(\theta)$ and integrated intensity $A_j(\theta)$ embedded our measured $I(\theta,\omega)$, a typical approach would be to fit the spectra $I_m(\omega)=I(\theta_m,\omega)$ for each measurement angle $\theta_m$ with a sum of Lorentzian peaks, that would provide the central frequency, linewidth and amplitude of each mode resonances. But the theory (cf. Eq.~(29) of the main text) shows that the spectra have a more complex, non-lorentzian shape, because of the steep frequency dependence of the thermal phonon distribution $n_T(\omega)$ and of the phonon-exciton interaction rate $\gamma_{xp}(\omega)$. These factors result in asymmetric peaks, such that using Lorentzian peaks is not good enough to extract our observables with error bars small enough to draw meaningful conclusions. Moreover the complex shape of the spectra often prevents the convergence of the fitting algorithm.

We thus take a different approach based on Eq.~(29) in order to fit the spectra $I_m(\omega)$ for each $\theta_m$, as it is found to provide the correct lineshape to a very good approximation, and an excellent first guess to provide to the fitting algorithm, using just the nominal experimental parameters. Since the goal is to extract the experimental observables independently of the theoretical predictions, we must add degrees of freedom to E Eq.~(29), such that the resonances of each mode are free to occur at any frequencies, linewidth and amplitude.

We thus add three degrees of freedom to Eq.~(29): $\delta_{Ec0}(\theta)$) which is added to the bare cavity dispersion relation in order to allow the cavity resonance to occur at any frequency, $A_0(\theta)$ which is a global multiplicative factor that allow capturing the experimental peak amplitudes exactly, and $\alpha(\theta)$ which is a multiplicative factor added to the ghost resonance such that its amplitude is free to take any value regardless of the theoretical one. Owing to its weak amplitude as compared to the normal mode, the ghost mode is also fitted separately from the normal one. The spectra $I_m(\omega)$ are then automatically fitted using a Levenberg-Marquardt fitting algorithm, in which these three variables are the fit outcome, plus a fourth one $B_e(\theta)$ that quantifies the intensity of the quasi-resonant extra peak.

The other parameters are determined before this fitting session in a separate way: the $k$-dependent excitonic linewidth $\gamma_x(\theta)=\gamma_{x,0}+\alpha k^2$ (involving two fit parameters) affects only the spectral width of the resonances, and is independent of the other fit parameters; it is thus determined  \textit{by hand} using a sub selection of spectra. The temperature $T$ and the quantum well cutoff momentum $q_{z,{\rm cut}}$ have a very similar influence on the intensity as a function of $\theta$ $A_j(\theta)$. But fortunately, at very low laser drive, we can be confident that the phonon temperature matches that of the cryostat. We thus fixed $T=T_c$ in the simulation and determined a single $q_{z,{\rm cut}}$ for all the measured temperatures ($T_c=[6.6,7.6,8.7,10,11,12]\,$K) that provides the best fit in terms of their measured $A_j(\theta)[T_c]$, in the very low laser drive regime, as is shown for $T_c=6.6$K and $T_c=12$K in Fig.\ref{fA2}.(a,b).

The cavity parameters such as the Rabi splitting $\Omega$, the bare exciton energy $\omega_x$ and the effective cavity mode index $n_{\rm cav}$ are also determined independently from the other parameters thanks to the contribution of the upper polariton branch in the measured spectra and dispersion relation. The UP resonance is also key in separating the contribution of $g_x$ and $g_s$ in the interacting regime as explained in the main text. 

Let us  point out that an additional emission peak is present in all the measured spectral functions at a fixed energy, quasi-resonant with the laser, and whose linewidth matches that of the polariton state at this energy (this extra peak is visible both in the experimental and theoretical $I(\theta,\omega)$ underneath the spectral filter in Fig.2 of the main text). This peak is not accounted for by the model, but its quasi-resonant character suggests that it results from the remaining spatial inhomogeneities in the system, such as weak in-plane disorder. The latter can indeed modulate the condensate density, and hence could create currents allowing low energy polariton to escape the laser-driven condensate mode without requiring the assistance of phonons. 

In practice, this quasi-resonant extra peak is added phenomenologically to Eq.~(29) as a third lorentzian peak with a $\theta-$independent energy and linewidth. It has a free amplitude $B_e(\theta)$, but a fixed width $\gamma_e$, and a fixed frequency $\omega_e$ very close to that of the laser. It contributes to the measured spectra via its low and high energy tails. Yet, as $\gamma_e$ and $\omega_e$ are only estimated from these spectral tails, the automatic fitting procedure described above is ran several times with slightly varying $\gamma_e$ and $\omega_e$ (within realistic ranges) and the resulting fits are analyzed quantitatively. The analysis provides the best fit to $\gamma_e$ and $\omega_e$, as well as their influence on the fitted variables. This influence is quantified via the goodness-of-fit-weighted standard deviation of the observables on $\omega_e$ and $\gamma_e$ and is added to the confidence intervals.

Note that $\Omega$, $n_{\rm cav}$, $q_{z, {\rm cut}}$, $\gamma_{\rm cav}$, as well as the laser frequency $\omega_l$, are parameters independent from both temperature and laser intensity (in the interacting regime): they are thus fixed once and for all. On the other hand, $\omega_x$ and to a lesser extent $\omega_{\rm cav}$ and the two parameters involved in $\gamma_x(\theta)$ depend on temperature and hence also on laser intensity.

\subsection{Comparison between theory and experiment}

The analysis presented above allows us to determine the best fit between the theoretical and experimental observables in the following way: in the theory, the added parameter $\delta_{Ec0}$ must be constant, such that its average value $\langle \delta_{Ec0}(\theta)\rangle$ provides a best fit of the bare cavity offset $E_{\rm cav,fit,0}=E_{\rm cav,fit}(\theta=0)=E_{\rm cav,guess,0}+\langle \delta_{Ec0}\rangle_\theta$, and its standard deviation over the angles provides a good estimate of the $1\sigma$ confidence interval of $E_{\rm cav,fit,0}$. The same argument is used to determine $A_{0,\rm fit}$ and its confidence interval. Finally, $\alpha_{\rm fit}=1$ is enforced as it is not free to take any other value in the theory. The other parameters are kept as determined by the experimental observable analysis described above. Fig.~\ref{fA2} shows selected raw spectra $I_m(\omega)$ in different conditions, as well as the theory best fit obtained using the method above.

Fig.~\ref{fA2}.(a,c) (Fig.~\ref{fA2}.(b,d)) are measurements realized at $T_c=6.6\,$K ($T_c=12\,$K). Fig.~\ref{fA2}.(a,b) (Fig.~\ref{fA2}.(c,d)) are obtained in the non-interacting (interacting) case. In the non-interacting case, the phonon temperature $T$ is taken in the theory as equal to that of the microcavity ($T_c$), while it is higher in the interacting case due to residual laser absorption: $T=15\,$K for $T_c=6.6\,$K, and $T=20\,$K for $T_c=12\,$K.

\subsection{Measured and calculated spectral functions at $T_c=6.6\,$K including intermediate laser powers}

Fig.2.(a,c) in the main text show the spectral functions for the two points labelled '1' and '3' in Fig.\ref{fA1}. Fig.\ref{fA3} show all the measured states '1' to '5'. state '2' is a high laser intensity but lower state of the bistable regime, showing a small but measurable amount of interactions energy $g_sn_x=0.03\,$meV. Points '4' and '5' are high interaction energy states situated beyond the bistable window, and characterized by $g_sn_x=0.210\,$meV and $g_sn_x=0.215\,$meV respectively. States '1' and '3' are discussed in the main text: they corresponds to interaction energies of $g_sn_x=0.0\,$meV and $g_sn_x=0.190\,$meV respectively.

The corresponding calculated thermal-phonons spectral functions $I_p(k,\omega)$ are shown in Fig.\ref{fA3}.(f-j). The phonon bath temperature is taken as $T=6.6\,$K in states '1' and '2' Fig.\ref{fA3}.(f,g) and $T=15\,$K in Fig.\ref{fA3}.(h-j). The excitonic and cavity photon transitions are found slightly redshifted by residual absorption induced temperature increase, namely by $\delta E_x=[0,0,10,30,50]\,\mu$eV and $\delta E_c=[0,0,0,30,30]\,\mu$eV.
\end{document}